\documentclass[prd,12pt]{article}

\usepackage{amsmath}
\usepackage{amssymb}

\usepackage{tikz}
\usepackage{amsmath,amssymb,graphicx,multirow,xspace,slashed,array,booktabs}
\usepackage[colorlinks=true,urlcolor=blue,anchorcolor=blue,citecolor=blue,filecolor=blue,linkcolor=blue,menucolor=blue]{hyperref}
\usepackage[compress,numbers]{natbib}
\usepackage{subfigure, placeins}
\usepackage{booktabs}
\usepackage{blindtext, rotating}
\usepackage{afterpage}
\usepackage{enumitem}
\usepackage{marvosym}
\usepackage{authblk}
\usepackage{verbatim}
\usepackage{multirow}
\usepackage{tikz-feynman}
\tikzfeynmanset{compat=1.1.0}
\usepackage{soul} 
\usepackage[normalem]{ulem}
\usepackage{pifont}
\usepackage{booktabs}

\usepackage{floatrow}
\newfloatcommand{capbtabbox}{table}[][\FBwidth]

\usepackage[font=footnotesize,labelfont=bf]{caption}

\usepackage{lineno}

\allowdisplaybreaks

\long\def\symbolfootnote[#1]#2{\begingroup%
\def\thefootnote{\fnsymbol{footnote}}\footnote[#1]{#2}\endgroup}

\allowdisplaybreaks

\makeatletter
	
	\@addtoreset{equation}{section}
\makeatother

\newcommand{\newc}{\newcommand}
\newc{\gsim}{\lower.7ex\hbox{$\;\stackrel{\textstyle>}{\sim}\;$}}
\newc{\lsim}{\lower.7ex\hbox{$\;\stackrel{\textstyle<}{\sim}\;$}}
\newc{\gev}{\,{\rm GeV}}
\newc{\mev}{\,{\rm MeV}}
\newc{\ev}{\,{\rm eV}}
\newc{\kev}{\,{\rm keV}}
\newc{\tev}{\,{\rm TeV}}
\newc{\MHT}{$H_T^{\text{miss}}$}
\newc{\MET}{$\slashed{E}_T$}
\newc{\MTT}{$M_{T2}$}

\def\ln{\mathop{\rm ln}}

\newc{\mz}{M_Z}
\newc{\mpl}{M_*}
\newc{\mw}{m_{\rm weak}}
\newc{\nr}[1]{N^c_R{}_{#1}}

\def\beq{\begin{equation}}
\def\eeq{\end{equation}}
\newcommand{\bea}{\begin{eqnarray}\begin{aligned}}
\newcommand{\eea}{\end{aligned}\end{eqnarray}}
\def\bitem{\begin{itemize}}
\def\eitem{\end{itemize}}
\newcommand{\ecm}{\, e \, {\rm cm}}

\begin{document}
\baselineskip 0.6cm

\begin{titlepage}

\vspace*{-0.0cm}

\thispagestyle{empty}

\begin{center}
\vskip 1cm

{\Large \bf 
Models for the Electric Dipole Moment and Anomalous Magnetic Moment of the Tau Lepton
}
\vskip 1cm

\vskip 1.0cm
{\large Yuichiro Nakai$^{1,2}$, Yoshihiro Shigekami$^{3}$, \\[1ex]
Peng Sun$^{1,2}$ and Zhihao Zhang$^{1,2}$}
\vskip 1.0cm
{\it
$^1$Tsung-Dao Lee Institute, Shanghai Jiao Tong University, \\ No.~1 Lisuo Road, Pudong New Area, Shanghai 201210, China \\
$^2$School of Physics and Astronomy, Shanghai Jiao Tong University, \\ 800 Dongchuan Road, Shanghai 200240, China \\
$^3$School of Physics, Henan Normal University, Xinxiang, Henan 453007, China}
\vskip 1.0cm

\end{center}

\vskip 1.0cm

\begin{abstract}

The Belle II experiment and other ongoing and projected lepton facilities are expected to greatly enhance the sensitivity to the electric dipole moment (EDM) and anomalous magnetic moment ($g-2$) for the tau lepton, making it timely to explore models that predict these observables. 
We present a class of models that generate a sizable EDM and $g-2$ of the tau lepton via radiative tau mass generation. 
Two benchmark models with different hypercharge assignments are investigated. 
The first model contains neutral fermions and charged scalars. 
We find that the model can predict a large signal of the tau EDM, $d_{\tau} = \mathcal{O}(10^{-19}) \ecm$, and $g-2$, $a_{\tau} = \mathcal{O}(10^{-5})$, which are within the reach of future updates of their measurements. 
In contrast, the second model, containing a charged fermion and neutral scalars, yields a similar magnitude for the $g-2$ but predicts a comparatively smaller EDM signal. 
Our models serve as benchmarks for new physics generating sizable EDM and $g-2$ of the tau lepton. 

\end{abstract}

\flushbottom

\end{titlepage}


\section{Introduction}
\label{intro}

Searches for the electric dipole moment (EDM) and anomalous magnetic moment ($g-2$) of charged leptons are driven by their key role as precision probes of fundamental laws of nature. 
These quantities are sensitive to quantum effects, making them powerful tools to test the Standard Model (SM) and to search for signs of new physics, complementing direct searches at colliders and offering insights into symmetries such as CP and lepton flavor. 
They are related to form factors in the effective photon-charged lepton coupling,
\begin{align}
\Gamma_{\ell}^{\alpha} (q^2) = F_1 (q^2) \gamma^{\alpha} + \left[ i F_2 (q^2) + F_3 (q^2) \gamma_5 \right] \frac{\sigma^{\alpha \beta} q_{\beta}}{2 m_{\ell}} \, ,
\end{align}
where $\ell = e, \mu, \tau$ indicates the charged lepton with mass $m_\ell$, $q^\nu$ is the four-momentum of the photon,
and $\sigma^{\alpha \beta} \equiv i [\gamma^{\alpha}, \gamma^{\beta}] / 2$. 
At the leading order, each form factor can be found as
\begin{align}
F_1 = 1 \, , \qquad F_2 = F_3 = 0 \, ,
\end{align}
while loop corrections and CP violation give non-zero results for $F_2$ and $F_3$. 
The $g-2$ and EDM are defined in terms of $F_2$ and $F_3$, respectively, with on-shell photon $q^2 = 0$:
\begin{align}
a_{\ell} \equiv \frac{g_{\ell} - 2}{2} = F_2 (0) \, , \qquad d_{\ell} = - \frac{e}{2 m_{\ell}} F_3 (0) \, ,
\end{align}
where $g_{\ell}$ is the $g$-factor of a charged lepton $\ell$, and $a_{\ell}$ receives the Schwinger correction at the one-loop level, $\alpha_{\rm em} / (2 \pi) \simeq 1.16 \times 10^{-3} \ecm$~\cite{Schwinger_PhysRev.73.416} with the fine-structure constant $\alpha_{\rm em} \simeq 1 / 137$. 
The SM contributions to the electron and muon $g-2$ are precisely calculated~\cite{Aliberti:2025beg} and compared with experiments
\cite{Fan:2022eto,Muong-2:2025xyk}. 
For the latest review on the muon $g-2$, see ref.~\cite{Athron:2025ets} and references therein. 
On the other hand, a nonzero EDM requires CP violation which arises predominantly from the phase in the Cabibbo-Kobayashi-Maskawa (CKM) matrix within the SM, resulting in extremely suppressed predictions for charged lepton EDMs. 
Therefore, the discovery of a nonzero EDM indicates the existence of physics beyond the SM. 
In fact, precise measurements of the electron EDM~\cite{ACME:2018yjb,Roussy:2022cmp} already put stringent constraints on a wide range of new physics models (see e.g. refs.~\cite{Nakai:2016atk,Cesarotti:2018huy,Ning:2025zfh} and references therein for implications on supersymmetry). 
While the current upper limit on the muon EDM~\cite{Muong-2:2008ebm,Ema:2021jds} hardly gives a constraint by itself, the projected experiments~\cite{Abe:2019thb,Adelmann:2021udj,Sakurai:2022tbk,muonEDMinitiative:2022fmk} will reach the sensitivity to explore new physics at the electroweak scale. 

The dipole moments of the tau lepton are currently much less constrained than those of the electron and muon due to the tau’s short lifetime, but they will provide a valuable window into new physics effects that scale with lepton mass. 
The SM contribution to the tau $g-2$ is precisely calculated and found as $a_{\tau}^{\rm SM} = (117717.1 \pm 3.9) \times 10^{-8}$, including updates of the hadronic vacuum polarization contributions~\cite{SMtaugm2EIDELMAN_2007,Keshavarzi:2019abf}. 
On the other hand, its measurements at the Large Hadron Collider (LHC) are still not as precise~\cite{ATLAS:2022ryk,CMS:2024qjo}:
\begin{align}
\text{ATLAS: }& -0.057 < a_{\tau} < 0.024 ~~~~~ (95 \% \: {\rm C.L.}) \, , \label{eq:MDMtauExpATLAS} \\[1ex]
\text{CMS: }& -0.0042 < a_{\tau} < 0.0062 ~~ (95 \% \: {\rm C.L.}) \, . \label{eq:MDMtauExpCMS}
\end{align}
The SM contribution to the tau EDM is tiny. 
At the quark level, the leading contribution is given at the four-loop level, $d_{\tau}^{\rm SM} = \mathcal{O}(10^{-47}) \ecm$~\cite{Pospelov:1991zt,Booth:1993af,Pospelov:2013sca}, while the hadron level long-distance effect enhances the contribution to $d_{\tau}^{\rm SM} \simeq -7.32 \times 10^{-38} \ecm$~\cite{tauEDM_Yamaguchi:2020eub,Yamaguchi:2020dsy}. 
The current experimental upper limits are~\cite{Roussy:2022cmp,Muong-2:2008ebm,CMS:2024qjo,Belle:2021ybo}
\begin{align}
-1.85 \times 10^{-17} \ecm &< {\rm Re} (d_{\tau}) < 0.61 \times 10^{-17} \ecm ~~ (95 \% \: {\rm C.L.}) \, , \label{eq:ExpRetauEDM} \\[1ex]
-1.03 \times 10^{-17} \ecm &< {\rm Im} (d_{\tau}) < 0.23 \times 10^{-17} \ecm ~~ (95 \% \: {\rm C.L.}) \, , \label{eq:ExpImtauEDM} \\[1ex]
|d_{\tau}| &< 2.9 \times 10^{-17} \ecm ~~ (95 \% \: {\rm C.L.}) \, . \label{eq:ExpImtauEDMCMS}
\end{align}
Note that the complex form of the limits is due to the off-shell photon in the $e^+ e^- \to \tau^+ \tau^-$ process. 
There is also an indirect limit on $|d_{\tau}|$ from the electron EDM constraint via the three-loop light-by-light mechanism~\cite{Ema:2022wxd}, which is
\begin{align}
|d_{\tau}| < \begin{cases}
1.1 \times 10^{-18} \ecm & \text{for $|d_e| < 1.1 \times 10^{-29} \ecm$~\cite{ACME:2018yjb} ,} \\[1ex]
4.1 \times 10^{-19} \ecm & \text{for $|d_e| < 4.1 \times 10^{-30} \ecm$~\cite{Roussy:2022cmp} ,}
\end{cases}
\label{eq:tauEDMindirect}
\end{align}
providing a stronger constraint compared with the direct bounds. 
Although measurements of the tau $g-2$ and EDM remain experimentally challenging, we can expect improved sensitivities for these observables by ongoing and projected experiments
such as the Belle II experiment~\cite{Belle-II:2010dht,Belle-II:2018jsg,Aihara:2024zds}, Beijing Electron-Positron Collider (BEPCII)~\cite{BESIII:2009fln,BESIII:2020nme,Bernreuther:2021uqm} and Circular Electron-Positron Collider (CEPC)~\cite{CEPCStudyGroup:2018ghi,CEPCPhysicsStudyGroup:2022uwl}. 
They will reach the sensitivities of $|a_{\tau}| \sim 10^{-5}$ and $|d_{\tau}| \sim 10^{-19} \ecm$~\cite{Belle2:2024tau,Crivellin:2021spu,USBelleIIGroup:2022qro}. 
Given such expected improvements in experimental sensitivity, it is natural to ask what kind of new physics models can be probed by measurements of the tau $g-2$ and EDM. 

New physics contributions to the dipole moments of a charged lepton are scaled by $1 / \Lambda_{\rm NP}^2$ with a new physics mass scale $\Lambda_{\rm NP}$, as the corresponding operators are effectively dimension-six, due to a Higgs field insertion for chirality flipping. 
Generally, the contributions are suppressed by a loop factor as well as the charged lepton mass. 
However, as discussed in ref.~\cite{Khaw:2022qxh}, we can avoid a loop suppression when the charged lepton mass is radiatively generated, because a loop factor is effectively absorbed into the mass in their expressions. 
In such a model to radiatively generate the tau lepton mass, by naive dimensional analysis, we can expect $a_{\tau} = \mathcal{O}(10^{-5})$ and $d_{\tau} = \mathcal{O}(10^{-19}) \ecm$ for $\Lambda_{\rm NP}$ at the electroweak scale. 
In the present paper, therefore, we investigate a class of radiatively generated tau mass models and their predictions for the tau $g-2$ and EDM.\footnote{Ref.~\cite{De:2023acg} presented a scalar leptoquark model for the tau $g-2$ and EDM. 
Collider searches for leptoquarks~\cite{CMS:2018oaj,CMS:2022nty,ATLAS:2021oiz} require their masses to be heavier than $\mathcal{O}(1)$ TeV, leading to smaller predictions for the tau dipole moments.}
Our models contain new exotic particles which exclusively couple to the tau lepton.\footnote{The radiative tau mass model has been discussed in ref.~\cite{Baker:2020vkh}, which explored constraints from the Higgs and electroweak measurements as well as dark matter phenomenology.}
A new source of CP violation arises from generally complex couplings whose phases cannot be completely removed by field redefinition. 
We explore two benchmark models with different hypercharge assignments of new particles. 

The rest of the paper is organized as follows. 
In section~\ref{sec:setup}, we describe our model setup and show detailed predictions for the tau $g-2$ and EDM. 
We start with a model-independent discussion and then consider concrete models. 
Section~\ref{sec:pheno} discusses relevant experimental and theoretical constraints. 
In section~\ref{sec:results}, we numerically explore viable parameter regions to give sizable tau $g-2$ and EDM which can be tested at future measurements. 
Section~\ref{sec:summary} is devoted to conclusions and discussions. 
Appendices summarize some formulae used in the main text.

\section{Model setup}
\label{sec:setup}

\begin{table}[!t]
\begin{center}
\begin{tabular}{|c||ccccccc|}
\hline
& $L_L^{\tau}$ & $\tau_R$ & $H$ & $\psi_L$ & $\psi_R$ & $\phi$ & $\eta$ \\ \hline \hline
$SU(2)_L$ & $\mathbf{2}$ & $\mathbf{1}$ & $\mathbf{2}$ & $\mathbf{1}$ & $\mathbf{1}$ & $\mathbf{2}$ & $\mathbf{1}$ \\
$Y$ & $- \frac{1}{2}$ & $-1$ & $\frac{1}{2}$ & $Y_{\psi}$ & $Y_{\psi}$ & $Y_{\psi} + \frac{1}{2}$ & $Y_{\psi} + 1$ \\ \hline
$L_{\tau}$ & $-$ & $-$ & $+$ & $+$ & $+$ & $-$ & $-$ \\
$X$ & $+$ & $+$ & $+$ & $-$ & $-$ & $-$ & $-$ \\ \hline
$S_a$ & $+$ & $-$ & $+$ & $+$ & $+$ & $+$ & $-$ \\ \hline
\end{tabular}
\caption{Charge assignments for the relevant particles in the radiative tau mass model. 
$L_L^{\tau}$ and $\tau_R$ represent the third generation of the left and right-handed leptons and $H$ is the SM Higgs field. 
$L_{\tau}$ and $X$ are $Z_2$ symmetries associated with the tau number and the exotic particle number, respectively, while $S_a$ is a softly broken $Z_2$ symmetry to forbid the tree-level tau Yukawa coupling.}
\label{tab:class1}
\end{center}
\end{table}

As discussed in the case of the radiative muon mass model~\cite{Baker:2021yli}, we consider new particles coupling to the tau lepton, whose charge assignments are summarized in Table~\ref{tab:class1}. 
The radiative mass for the tau lepton can be obtained by the following two Yukawa terms:
\begin{align}
\mathcal{L}_{\rm rad} = - y_{\phi} \overline{L_L^{\tau}} \phi^{\dagger} \psi_R - y_{\eta} \overline{\psi}_L \eta \tau_R + {\rm h.c.} \, ,
\end{align}
where these terms always exist in any choice of $Y_{\psi}$. 
Note that the two Yukawa couplings $y_{\phi, \eta}$ can be real and positive by using rephasing degrees of freedom. 
Here, we also need a mixing between $\phi$ and $\eta$, which is
\begin{align}
V_{\rm scl} \supset a H \eta^{\dagger} \phi + {\rm h.c.} \, , \label{eq:commonVscl}
\end{align}
with a mass dimension one parameter $a$ in the scalar potential. 
The radiative tau lepton mass can be then obtained as
\begin{align}
m_{\tau}^{\rm rad} = \frac{y_{\phi} y_{\eta}}{16 \pi^2} \mathcal{F}_{\rm mdl} \, ,
\label{eq:mtauradGen}
\end{align}
where $\mathcal{F}_{\rm mdl}$ is a model-dependent part with mass dimension 1 which is expected to be scaled by the mass of $\psi_{L, R}$, due to mass flipping from the inner $\psi_{L, R}$ line, and also depends on the mixing angles for exotic particles. 
We emphasize that $\mathcal{F}_{\rm mdl}$ is generally complex, originated from the mixing of $\psi_{L, R}$ or $\phi$ and $\eta$, which arrows us to parameterize it as $\mathcal{F}_{\rm mdl} = |\mathcal{F}_{\rm mdl}| e^{i \theta_{\tau}}$. 
The phase $\theta_{\tau}$ is equivalent to the phase of $m_{\tau}^{\rm rad}$ due to real values of $y_{\phi, \eta}$. 
The absolute value of $m_{\tau}^{\rm rad}$ should be the measured mass of the tau lepton, and hence, we have the following relation:
\begin{align}
\frac{y_{\phi} y_{\eta}}{16 \pi^2} = \frac{m_{\tau}^{\rm exp}}{|\mathcal{F}_{\rm mdl}|} \, ,
\label{eq:yphiyetarelation}
\end{align}
which is the main essence for generating large $g-2$ and EDM in the radiative mass model~\cite{Khaw:2022qxh,Baker:2020vkh,Baker:2021yli}. 
Note that compared to the case of the radiative muon mass model, the radiative tau mass model tends to predict larger Yukawa couplings, $y_{\phi}, y_{\eta}$. 
From Eq.~\eqref{eq:yphiyetarelation}, one can easily see that
\begin{align}
y_{\phi} y_{\eta} = 16 \pi^2 \frac{m_{\tau}^{\rm exp}}{|\mathcal{F}_{\rm mdl}|} \sim 10 \, ,
\label{eq:yphiyetanum}
\end{align}
due to the heavy mass of the tau lepton, $m_{\tau}^{\rm exp} = 1.77693$ GeV~\cite{ParticleDataGroup:2024cfk}. 
Here, we assume that the masses of exotic particles are $\mathcal{O} (100)$ GeV, and the exotic mixing angles give $1 / 4$-$1 / 3$ suppression factor on $|\mathcal{F}_{\rm mdl}|$, which becomes an enhancement factor on $y_{\phi} y_{\eta}$. 
Therefore, if we need to focus on a parameter space where the exotic mass scale is $\mathcal{O} (100)$ GeV to obtain large tau $g-2$ and/or EDM, the bound on $y_{\phi} y_{\eta} < 4 \pi$ will constrain the parameter space severely. 

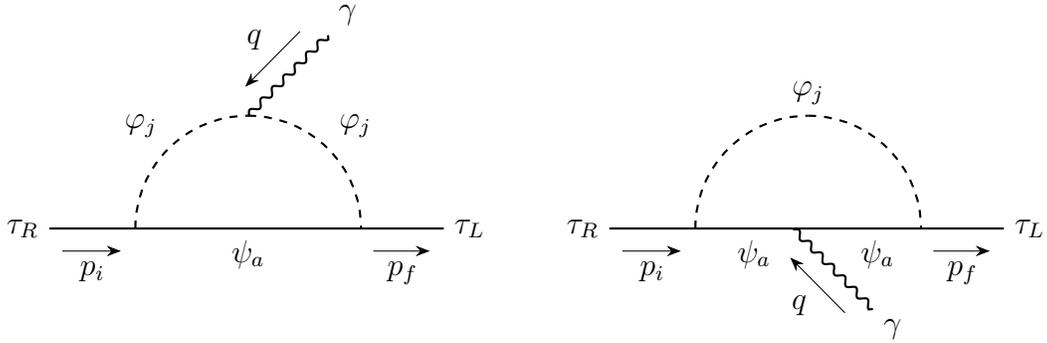
\begin{figure}[!t]
\begin{center}
\begin{tikzpicture}
\begin{feynman}[large]
\vertex (a1) {\(\tau_R\)};
\vertex [right=1.5cm of a1] (b1);
\vertex [right=3cm of b1] (c1);
\vertex [right=1.1cm of c1] (d1) {\(\tau_L\)};
\vertex [above right=2.12132cm of b1] (f1);
\vertex [above right=1.5cm of f1] (g1) {\(\gamma\)};
\diagram [medium] {
(a1) -- [momentum'=\(p_i\)] (b1) -- [edge label'=\(\psi_a\)] (c1) -- [momentum'=\(p_f\)] (d1),
(b1) -- [scalar, quarter left, edge label=\(\varphi_j\)] (f1) -- [scalar, quarter left, edge label=\(\varphi_j\)] (c1),
(f1) -- [photon, rmomentum=\(q\)] (g1),
};
\vertex [right=1.5cm of d1] (a3) {\(\tau_R\)};
\vertex [right=1.5cm of a3] (b3);
\vertex [right=1.3cm of b3] (f3);
\vertex [right=1.7cm of f3] (c3);
\vertex [right=1.1cm of c3] (d3) {\(\tau_L\)};
\vertex [above right=2.12132cm of b3] (h3);
\vertex [below right=1.5cm of f3] (g3) {\(\gamma\)};
\diagram [medium] {
(a3) -- [momentum'=\(p_i\)] (b3) -- [pos=0.6, edge label'=\(\psi_a\)] (f3) -- [pos=0.66, edge label'=\(\psi_a\)] (c3) -- [momentum'=\(p_f\)] (d3),
(b3) -- [scalar, quarter left, pos=1, edge label=\(\varphi_j\)] (h3) -- [scalar, quarter left] (c3),
(b3) -- [scalar, in=180, out=0] -- (c3),
(f3) -- [photon, rmomentum'=\(q\)] (g3),
};
\end{feynman}
\end{tikzpicture}
\end{center}
\vspace{-0.5cm}
\caption{Diagrams relevant to $C_T (q^2)$ and $C_{T'} (q^2)$. 
$q = p_f - p_i$ is the four-momentum of the photon, and $\varphi_j$ and $\psi_a$ are mass eigenstates for the exotic particles. }
\label{fig:dipolediagrams}
\end{figure}

The dipole operators are defined as
\begin{align}
\mathcal{L}_{\rm dipole} = - \frac{e}{2} C_T (q^2) \left( \bar{\tau} \sigma^{\alpha \beta} \tau \right) F_{\alpha \beta} - \frac{e}{2} C_{T'} (q^2) \left( \bar{\tau} i \sigma^{\alpha \beta} \gamma_5 \tau \right) F_{\alpha \beta} \, ,
\label{eq:dipoleOPsGen}
\end{align}
where $q$ and $F_{\alpha \beta}$ are the four-momentum and the field strength of the photon, respectively, and relevant diagrams to $C_T (q^2)$ and $C_{T'} (q^2)$ in the radiative tau mass model are shown in Fig.~\ref{fig:dipolediagrams}. 
The phase $\theta_{\tau}$ of the tau lepton mass can be removed by a chiral rotation,
\begin{align}
\tau \to e^{- i \theta_{\tau} \gamma_5 / 2} \tau \, , \label{eq:tauChiral}
\end{align}
which also affects the dipole operators. 
The final results for the tau $g-2$ and EDM are then given by
\begin{align}
\mathcal{L}_{\rm dipole} &= - \frac{e}{4 m_{\tau}} a_{\tau} \left( \bar{\tau} \sigma^{\alpha \beta} \tau \right) F_{\alpha \beta} - \frac{i}{2} d_{\tau} \left( \bar{\tau} \sigma^{\alpha \beta} \gamma_5 \tau \right) F_{\alpha \beta} \, , \label{eq:DipoleGen} \\[1ex]
a_{\tau} &= 2 m_{\tau} \left( C_T (0) \cos \theta_{\tau} + C_{T'} (0) \sin \theta_{\tau} \right) \, , \label{eq:atauGen} \\[1ex]
d_{\tau} &= e \left( C_{T'} (0) \cos \theta_{\tau} - C_T (0) \sin \theta_{\tau} \right) \, . \label{eq:dtauGen}
\end{align}
It is useful that $\sin \theta_{\tau}$ and $\cos \theta_{\tau}$ can be determined by $\mathcal{F}_{\rm mdl}$ as
\begin{align}
\sin \theta_{\tau} = \frac{{\rm Im} (\mathcal{F}_{\rm mdl})}{|\mathcal{F}_{\rm mdl}|} \, , \qquad \cos \theta_{\tau} = \frac{{\rm Re} (\mathcal{F}_{\rm mdl})}{|\mathcal{F}_{\rm mdl}|} \, . \label{eq:sincosthetatau}
\end{align}

To obtain only the desired terms in the model, we impose three discrete $Z_2$ symmetries: $L_{\tau}$, $X$ and $S_a$. 
The last one, $S_a$, plays a role to forbid the tree-level tau Yukawa coupling and is softly broken by the terms in Eq.~\eqref{eq:commonVscl}. 
On the other hand, $L_{\tau}$ and $X$ which are associated with the tau number and the exotic number, respectively, allow appropriate Yukawa terms with exotic particles and the tau lepton exclusively.
The $L_{\tau}$ symmetry forbids lepton-flavor violating (LFV) processes associated with $\tau$ decays, e.g., $\tau \to \ell_i \gamma$ and $\tau \to \ell_i \bar{\ell_j} \ell_k$ with $\ell_{i, j, k} = e, \mu$. 
Since the left-handed doublet lepton $L^{\tau}_L$ is charged under $L_{\tau}$, it is problematic to reproduce the observed neutrino mixing angles. 
Therefore, for the neutrino sector, we need an extension of the model, as discussed in ref.~\cite{Khaw:2022qxh}\footnote{Recently, the authors in ref.~\cite{Giarnetti:2025idu} explored models with global $U(2)_F$ symmetry in the context of proper neutrino observables as well as LFV and lepton anomalous magnetic moments.}. 
In the present paper, we expect that this extension does not give a relevant effect on our numerical analyses.

\subsection{Majorana fermion model}
\label{sec:MFmodel}

Let us now consider a concrete model for the radiative tau lepton mass. 
We choose the hypercharge $Y_{\psi}$ in Table~\ref{tab:class1} to be 0, and hence, we have SM singlet fermions, which can be Majorana. 
Then, we call it as the Majorana fermion (MF) model. 

The Lagrangian and scalar potential of the MF model are
\begin{align}
\mathcal{L}_{\rm MF} &\supset \left( - y_{\phi} \overline{L_L^{\tau}} \phi^{\dagger} \psi_R - y_{\eta} \overline{\psi}_L \eta \tau_R - m_D \overline{\psi}_L \psi_R - \frac{m_{LL}}{2} \overline{\psi}_L \psi^c_L - \frac{m_{RR}}{2} \overline{\psi^c_R} \psi_R + {\rm h.c.} \right) - V_{\rm scl, \, MF} \, , \label{eq:LagpartMF} \\
V_{\rm scl, \, MF} &= \sum_{s = H, \phi, \eta} \left[ m_s^2 s^{\dagger} s + \frac{\lambda_s}{2} (s^{\dagger} s)^2 \right] + \lambda_{H \phi} (H^{\dagger} H) (\phi^{\dagger} \phi) + \lambda_{H \eta} (H^{\dagger} H) (\eta^{\dagger} \eta) + \lambda_{\phi \eta} (\phi^{\dagger} \phi) (\eta^{\dagger} \eta) \nonumber \\
&\hspace{1.2em} + \lambda'_{H \phi} (H^{\dagger} \phi) (\phi^{\dagger} H) + \left( a H \eta^{\dagger} \phi + \frac{\lambda''_{H \phi}}{2} (H^{\dagger} \phi)^2 + {\rm h.c.} \right) \, . \label{eq:VsclMF}
\end{align}
Due to rephasing degrees of freedom, all couplings except for one of $m_D, m_{LL}, m_{RR}$ can be real and positive. 
The physical phase of the MF model is then
\begin{align}
\theta_{\rm phys}^{\rm MF} \equiv \frac{1}{2} \arg \left( \frac{m_{LL} m_{RR}}{m_D^2} \right) = \frac{1}{2} \Bigl( \theta_L + \theta_R - 2 \theta_D \Bigr) \, ,
\label{eq:CPphysMF}
\end{align}
where $\theta_{D, L, R}$ are phases of $m_D$, $m_{LL}$, and $m_{RR}$, respectively. 
Note that the left and right-handed Majorana fermions can be defined as $\psi_{L, R}^M \equiv \psi_{L, R} + (\psi_{L, R})^c$. 
We assume that two exotic scalars $\phi$ and $\eta$ have zero vacuum expectation values (VEVs). 
This leads to no mixing between $H$ and $\phi, \eta$, and hence, all scalars in the MF model can be parameterized as
\begin{align}
H = \begin{pmatrix}
G^+ \\
\frac{1}{\sqrt{2}} \left( v_H + h^0 + i G^0 \right)
\end{pmatrix} \, , \quad \phi = \begin{pmatrix}
\phi^+ \\
\frac{1}{\sqrt{2}} \left( \sigma_{\phi} + i a_{\phi} \right)
\end{pmatrix} \, , \qquad \eta = \eta^+ \, ,
\label{eq:scalarinMF}
\end{align}
where $v_H = 246.22$ GeV is the SM Higgs VEV, $G^+$ and $G^0$ are Nambu-Goldstone modes, and $h^0$ is the SM Higgs boson. 

In order to obtain the mass eigenvalues for all exotic particles, we need to diagonalize the corresponding mass matrices for $(\psi_L, \psi_R)$, $(\sigma_{\phi}, a_{\phi})$ and $(\phi^+, \eta^+)$. 
The details are discussed in ref.~\cite{Khaw:2022qxh}, which shows
\begin{align}
m_{\psi_1}^2 &= \frac{1}{2} \Bigl( m_{LL}^2 + m_{RR}^2 + 2 \left| m_D \right|^2 - \Delta m_{\psi}^2 \Bigr) \, , \label{eq:eigenpsi1} \\[1ex]
m_{\psi_2}^2 &= \frac{1}{2} \Bigl( m_{LL}^2 + m_{RR}^2 + 2 \left| m_D \right|^2 + \Delta m_{\psi}^2 \Bigr) \, , \label{eq:eigenpsi2} \\[1ex]
\Delta m_{\psi}^2 &= \sqrt{\left( m_{LL}^2 - m_{RR}^2 \right)^2 + 4 \left| m_D \right|^2 \Bigl| m_{LL} e^{- i \theta_{\rm phys}^{\rm MF}} + m_{RR} e^{i \theta_{\rm phys}^{\rm MF}} \Bigr|^2} \, , \label{eq:defDelmpsi2} \\[1ex]
\sin 2 \alpha &= \frac{2 | m_D |}{\Delta m_{\psi}^2} \Bigl| m_{LL} e^{- i \theta_{\rm phys}^{\rm MF}} + m_{RR} e^{i \theta_{\rm phys}^{\rm MF}} \Bigr| \, , \\[1ex]
\tan \theta_{\rm MF} &= - \frac{m_{LL} - m_{RR}}{m_{LL} + m_{RR}} \tan \theta_{\rm phys}^{\rm MF} \, , \label{eq:tantau}
\end{align}
for the Majorana fermion sector and
\begin{align}
m_{\varphi_1^+}^2 &= \frac{1}{2} \left[ M_{\phi}^2 + M_{\eta}^2 - \sqrt{(M_{\phi}^2 - M_{\eta}^2)^2 + 2 a^2 v_H^2} \right] \, , \label{eq:eigenpm1} \\[1ex]
m_{\varphi_2^+}^2 &= \frac{1}{2} \left[ M_{\phi}^2 + M_{\eta}^2 + \sqrt{(M_{\phi}^2 - M_{\eta}^2)^2 + 2 a^2 v_H^2} \right] \, , \label{eq:eigenpm2} \\[1ex]
\sin 2 \theta &= \frac{\sqrt{2} a v_H}{m_{\varphi_2^+}^2 - m_{\varphi_1^+}^2} \, , \label{eq:mixanglepm}
\end{align}
for the exotic charged scalar sector, with $M_{\phi, \eta}^2 \equiv m_{\phi, \eta}^2 + \frac{\lambda_{H \phi, H \eta}}{2} v_H^2$. 
The relations between flavor states and mass eigenstates are given by
\begin{align}
\psi_L &= \psi_1^c c_{\alpha} + \psi_2^c s_{\alpha} e^{- i \theta_{\rm MF}} \, , &&\psi_R = - \psi_1 s_{\alpha} e^{- i \theta_{\rm MF}} + \psi_2 c_{\alpha} \, ,
\label{eq:MajoranamasseigenMF} \\[1ex]
\phi^{\pm} &= \varphi_1^{\pm} c_{\theta} + \varphi_2^{\pm} s_{\theta} \, , &&\eta^{\pm} = - \varphi_1^{\pm} s_{\theta} + \varphi_2^{\pm} c_{\theta} \, . \label{eq:scalarmasseigenMF}
\end{align}
Note that the exotic neutral scalars do not mix with the neutral component of $H$, and moreover, $\sigma_{\phi}$ and $a_{\phi}$ also do not mix with each other, due to no CP violation in the scalar potential. 
Therefore, their mass eigenvalues can be found as
\begin{align}
m_{\sigma_{\phi}}^2 = M_{\phi}^2 + \frac{\lambda'_{H \phi} + \lambda''_{H \phi}}{2} v_H^2 \, , \qquad m_{a_{\phi}}^2 = M_{\phi}^2 + \frac{\lambda'_{H \phi} - \lambda''_{H \phi}}{2} v_H^2 \, .
\end{align}

By substituting the mass eigenstates into the original Lagrangian \eqref{eq:LagpartMF}, the relevant terms for calculating the radiative mass and effective Yukawa coupling as well as the dipole operators of the tau lepton are
\begin{align}
\mathcal{L}_{\rm MF} \supset \Bigl( - y_L^{j a} \bar{\tau}_L \varphi_j^- \psi_a - y_R^{j a} \overline{\psi_a^c} \varphi_j^+ \tau_R + {\rm h.c.} \Bigr) - \frac{A_{j k}}{\sqrt{2}} h^0 \varphi_j^- \varphi_k^+ \, ,
\label{eq:relevanttermsMF}
\end{align}
with the explicit forms of $y_{L, R}^{j a}$ and $A_{j k}$ in Table~\ref{tab:yLRAjk}. 
\begin{table}[!t]
\begin{center}
\begin{tabular}{|c|cc||c|c|}
\hline
$(j, a)$ & $y_L^{j a}$ & $y_R^{j a}$ & $(j, k)$ & $A_{j k}$ \\ \hline
$(1, 1)$ & $- y_{\phi} c_{\theta} s_{\alpha} e^{- i \theta_{\rm MF}}$ & $- y_{\eta} s_{\theta} c_{\alpha}$                & $(1, 1)$ & $- a s_{2 \theta} + \sqrt{2} v_H \left( \lambda_{H \phi} c_{\theta}^2 + \lambda_{H \eta} s_{\theta}^2 \right)$ \\
$(1, 2)$ & $y_{\phi} c_{\theta} c_{\alpha}$                    & $- y_{\eta} s_{\theta} s_{\alpha} e^{i \theta_{\rm MF}}$ & $(1, 2)$ & $a c_{2 \theta} + \sqrt{2} v_H \left( \lambda_{H \phi} - \lambda_{H \eta} \right) s_{\theta} c_{\theta}$ \\
$(2, 1)$ & $- y_{\phi} s_{\theta} s_{\alpha} e^{- i \theta_{\rm MF}}$ & $y_{\eta} c_{\theta} c_{\alpha}$                  & $(2, 1)$ & $a c_{2 \theta} + \sqrt{2} v_H \left( \lambda_{H \phi} - \lambda_{H \eta} \right) s_{\theta} c_{\theta}$ \\
$(2, 2)$ & $y_{\phi} s_{\theta} c_{\alpha}$                    & $y_{\eta} c_{\theta} s_{\alpha} e^{i \theta_{\rm MF}}$   & $(2, 2)$ & $a s_{2 \theta} + \sqrt{2} v_H \left( \lambda_{H \phi} s_{\theta}^2 + \lambda_{H \eta} c_{\theta}^2 \right)$ \\[0.3ex] \hline
\end{tabular}
\caption{Yukawa couplings for the tau lepton and exotic particles (left block) and scalar trilinear couplings (right block) in Eq.~\eqref{eq:relevanttermsMF}.}
\label{tab:yLRAjk}
\end{center}
\end{table}
The radiative mass and effective Yukawa coupling of the tau lepton in the MF model can be obtained by replacing $\mu \to \tau$ in the corresponding expressions of ref.~\cite{Khaw:2022qxh},
\begin{align}
\mathcal{L}_{\rm eff} &\supset - m_{\tau}^{\rm rad} \bar{\tau}_L \tau_R - \frac{y_{\tau}^{\rm eff}}{\sqrt{2}} \bar{\tau}_L \tau_R h^0 + {\rm h.c.} \, , \\[1.2ex]
m_{\tau}^{\rm rad} &= \sum_{j, a} \frac{y_L^{j a} y_R^{j a}}{16 \pi^2} m_{\psi_a} \left( \frac{m_{\psi_a}^2 B_0 (0, m_{\psi_a}^2, m_{\psi_a}^2) - m_{\varphi_j^+}^2 B_0 (0, m_{\varphi_j^+}^2, m_{\varphi_j^+}^2)}{m_{\varphi_j^+}^2 - m_{\psi_a}^2} - 1 \right) \nonumber \\[0.6ex]
&= \frac{y_{\phi} y_{\eta}}{16 \pi^2} \frac{s_{2 \theta} s_{2 \alpha}}{4} \mathcal{F}_{\rm MF} (x_{1, 1}, x_{1, 2}, x_{2, 1}, x_{2, 2}) \, , \label{eq:radiativemtauMF} \\[1.0ex]
y_{\tau}^{\rm eff} (p_{h^0}^2) &= - \sum_{j, k, a} \frac{y_L^{j a} y_R^{k a} A_{j k}}{16 \pi^2} m_{\psi_a} C_0 (m_{\tau}^2, p_{h^0}^2, m_{\tau}^2, m_{\psi_a}^2, m_{\varphi_j^+}^2, m_{\varphi_k^+}^2) \, . \label{eq:YefftauMF}
\end{align}
Here $p_{h^0}$ is the four-momentum of the SM Higgs boson, $B_0$ and $C_0$ are loop integral functions defined in appendix~\ref{app:PVint}, and
\begin{align}
\mathcal{F}_{\rm MF} (x_{1, 1}, x_{1, 2}, x_{2, 1}, x_{2, 2}) &\equiv m_{\psi_1} e^{- i \theta_{\rm MF}} \Bigl[ I_{\rm rad} (x_{1, 1}) - I_{\rm rad} (x_{2, 1}) \Bigr] \nonumber \\[0.3ex]
&\hspace{2.4em}- m_{\psi_2} e^{i \theta_{\rm MF}} \Bigl[ I_{\rm rad} (x_{1, 2}) - I_{\rm rad} (x_{2, 2}) \Bigr] \, , \label{eq:FfuncMF}
\end{align}
with $x_{j, a} \equiv m_{\varphi_j^+}^2 / m_{\psi_a}^2$ and
\begin{align}
I_{\rm rad} (x) \equiv \frac{x}{x - 1} \ln x \, . \label{eq:IradDef}
\end{align}
As expected in Eq.~\eqref{eq:mtauradGen}, $\mathcal{F}_{\rm MF} (x_{1, 1}, x_{1, 2}, x_{2, 1}, x_{2, 2})$ is scaled by $m_{\psi_{1, 2}}$. 
From Eq.~\eqref{eq:radiativemtauMF}, the size of $y_{\phi} y_{\eta}$ depends not only on the scale of $m_{\psi_a}$ but also on the mixing angles for the Majorana fermion and exotic charged scalar sectors. 
Therefore, $s_{2 \alpha} \sim s_{2 \theta} \sim 1$ and $|\mathcal{F}_{\rm MF} (x_{1, 1}, x_{1, 2}, x_{2, 1}, x_{2, 2})| \gtrsim \mathcal{O} (100)$ GeV are required. 

In the MF model, $C_T (q^2)$ and $C_{T'} (q^2)$ can be estimated by the left diagram of Fig.~\ref{fig:dipolediagrams},
\begin{align}
C_T^{\rm MF} (0) &= \sum_{j, a} \frac{{\rm Re}[y_L^{j a} y_R^{j a}]}{16 \pi^2} m_{\psi_a} \Bigl[ C_0 (m_{\psi_a}^2, m_{\varphi_j^+}^2) + 2 C_1 (m_{\psi_a}^2, m_{\varphi_j^+}^2) \Bigr] \, , \label{eq:CT0MF} \\[1ex]
C_{T'}^{\rm MF} (0) &= \sum_{j, a} \frac{{\rm Im}[y_L^{j a} y_R^{j a}]}{16 \pi^2} m_{\psi_a} \Bigl[ C_0 (m_{\psi_a}^2, m_{\varphi_j^+}^2) + 2 C_1 (m_{\psi_a}^2, m_{\varphi_j^+}^2) \Bigr] \, , \label{eq:CTp0MF}
\end{align}
where
\begin{align}
C_0 (m_{\psi_a}^2, m_{\varphi_j^+}^2) &\equiv C_0 (m_{\tau}^2, 0, m_{\tau}^2, m_{\psi_a}^2, m_{\varphi_j^+}^2, m_{\varphi_j^+}^2) \approx C_0 (0, 0, 0, m_{\psi_a}^2, m_{\varphi_j^+}^2, m_{\varphi_j^+}^2) \, , \\
C_1 (m_{\psi_a}^2, m_{\varphi_j^+}^2) &\equiv C_1 (m_{\tau}^2, 0, m_{\tau}^2, m_{\psi_a}^2, m_{\varphi_j^+}^2, m_{\varphi_j^+}^2) \approx C_1 (0, 0, 0, m_{\psi_a}^2, m_{\varphi_j^+}^2, m_{\varphi_j^+}^2) \, ,
\end{align}
are loop integral functions whose definitions are summarized in appendix~\ref{app:PVint}, and the last approximations can be applied when $m_{\tau}^2 \ll m_{\varphi_j^+}^2, m_{\psi_a}^2$. 
In this limit, we find the following simple analytical form:
\begin{align}
C_0 (m_{\psi_a}^2, m_{\varphi_j^+}^2) + 2 C_1 (m_{\psi_a}^2, m_{\varphi_j^+}^2) \approx \frac{1}{m_{\psi_a}^2} \frac{x_{j, a}^2 - 1 - 2 x_{j, a} \ln x_{j, a}}{2 (1 - x_{j, a})^3} \, .
\end{align}

\subsection{Real scalar model}
\label{sec:RSmodel}

As an another explicit model, we choose the hypercharge to be $Y_{\psi} = -1$ and follow the same steps as those of the MF model. 
In this case, we have one real singlet scalar, and therefore, we call this model as the real scalar (RS) model. 

The Lagrangian and scalar potential are similar to those of the MF model,
\begin{align}
\mathcal{L}_{\rm RS} &\supset \left( - y_{\phi} \overline{L_L^{\tau}} \phi^{\dagger} \psi_R - y_{\eta} \overline{\psi_L} \eta \tau_R - m_{\psi} \overline{\psi_L} \psi_R + {\rm h.c.} \right) - V_{\rm scl, \, RS} \, , \label{eq:LagpartRS} \\[1ex]
V_{\rm scl, \, RS} &= \sum_{s = H, \phi, \eta} \left[ m_s^2 s^{\dagger} s + \frac{\lambda_s}{2} (s^{\dagger} s)^2 \right] + \lambda_{H \phi} (H^{\dagger} H) (\phi^{\dagger} \phi) + \lambda_{H \eta} (H^{\dagger} H) \eta^2 + \lambda_{\phi \eta} (\phi^{\dagger} \phi) \eta^2 \nonumber \\
&\hspace{1.2em} + \lambda'_{H \phi} (H^{\dagger} \phi) (\phi^{\dagger} H) + \left( - a H \eta \phi + \frac{\lambda''_{H \phi}}{2} (H \phi)^2 + {\rm h.c.} \right) \, . \label{eq:VsclRS}
\end{align}
In this model, $\eta$ is the real scalar and hence, one of phases of $a$ and $\lambda''_{H \phi}$ cannot be removed, which leads to the physical CP phase of the RS model,
\begin{align}
\theta_{\rm phys}^{\rm RS} = \arg \left( \frac{\lambda''_{H \phi}}{a^2} \right) = \theta_{\lambda''_{H \phi}} - 2 \theta_a \, ,
\label{eq:CPphysRS}
\end{align}
where $\theta_{\lambda''_{H\phi}}$ and $\theta_a$ are the phases of $\lambda''_{H\phi}$ and $a$, respectively. 

As in the case of the MF model, we assume that $\phi$ and $\eta$ do not acquire nonzero VEVs, leading to the following parameterizations for doublet scalar fields:
\begin{align}
H = \begin{pmatrix}
G^+ \\
\frac{1}{\sqrt{2}} \left( v_H + h^0 + i G^0 \right)
\end{pmatrix} \, , \qquad \phi = \begin{pmatrix}
\frac{1}{\sqrt{2}} \left( \sigma_{\phi} + i a_{\phi} \right) \\
\phi^-
\end{pmatrix} \, ,
\label{eq:scalarinRS}
\end{align}
where $G^{\pm}$ and $h^0, G^0$ do not mix with $\phi^{\pm}$ and $\sigma_{\phi}, a_{\phi}, \eta$, respectively. 
We obtain the mass for $\phi^{\pm}$ and the mass-squared matrix for exotic neutral scalars in the basis of $(\sigma_{\phi}, \eta, a_{\phi})$,
\begin{align}
m_{\phi^{\pm}}^2 &= m_{\phi}^2 + \frac{\lambda_{H \phi} + \lambda'_{H \phi}}{2} v_H^2 \, , \label{eq:mphipm} \\[1ex]
\mathcal{M}_0^2 &= \begin{pmatrix}
m_{\phi}^2 + \frac{\bar{\lambda}_+}{2} v_H^2 & {\rm Re}(a) v_H & - \frac{{\rm Im}(\lambda''_{H \phi})}{2} v_H^2 \\
{\rm Re}(a) v_H & 2 m_{\eta}^2 + \lambda_{H \eta} v_H^2 & - {\rm Im}(a) v_H \\
- \frac{{\rm Im}(\lambda''_{H \phi})}{2} v_H^2 & - {\rm Im}(a) v_H & m_{\phi}^2 + \frac{\bar{\lambda}_-}{2} v_H^2
\end{pmatrix} \, , \label{eq:massmatrix}
\end{align}
where $\bar{\lambda}_{\pm} \equiv \lambda_{H \phi} \pm {\rm Re}(\lambda''_{H \phi})$. 
By using a $3 \times 3$ orthogonal matrix $R$, we can diagonalize $\mathcal{M}_0^2$ as
\begin{align}
R^T \mathcal{M}_0^2 R = {\rm diag}(m_{\varphi_1}^2, m_{\varphi_2}^2, m_{\varphi_3}^2) \, ,
\end{align}
and the flavor eigenstates can be written in terms of the mass eigenstates $\varphi_i$ as
\begin{align}
\sigma_{\phi} = \sum_{j = 1}^3 R_{1 j} \varphi_j \, , \quad \eta = \sum_{j = 1}^3 R_{2 j} \varphi_j \, , \quad a_{\phi} = \sum_{j = 1}^3 R_{3 j} \varphi_j \, .
\end{align}
Note that the assumption of VEVs for $\phi$ and $\eta$ results in no mixing between $\tau$ and $\psi$, and hence, the mass of $\psi$ is solely parameterized by $m_{\psi}$ in Eq.~\eqref{eq:LagpartRS}. 

By using the mass eigenstates, we obtain Yukawa couplings of the tau lepton to $\varphi_i$ and scalar trilinear couplings. 
We write the relevant terms as
\begin{align}
\mathcal{L}_{\rm RS} \supset \Bigl( - \frac{y_L^{(j)}}{\sqrt{2}} \overline{\tau_L} \varphi_j \psi_R - y_R^{(j)} \overline{\psi_L} \varphi_j \tau_R + {\rm h.c.} \Bigr) - A_{j k} h^0 \varphi_j \varphi_k \, ,
\label{eq:relevantterms}
\end{align}
where $y_L^{(j)} \equiv y_{\phi} \left( R_{1 j} - i R_{3 j} \right)$ and $y_R^{(j)} \equiv y_{\eta} R_{2 j}$, and $A_{j k}$ is summarized in Table~\ref{tab:Aij}. 
\begin{table}[!t]
\centering
\begin{tabular}{|c|c|}
\hline
\multirow{2}{*}{$A_{j j}$} & ${\displaystyle \frac{1}{2} \left( \bar{\lambda}_+ R_{1 j}^2 + \bar{\lambda}_- R_{3 j}^2 + 2 \lambda_{H \eta} R_{2 j}^2 \right) v_H + R_{1 j} R_{2 j} {\rm Re}(a)} $\\
$ $ & $ {- R_{2 j} R_{3 j} {\rm Im}(a) - {\rm Im}(\lambda''_{H \phi}) R_{1 j} R_{3 j} v_H}$ \\[0.3ex]
\multirow{2}{*}{$A_{j k}$} & ${\displaystyle \left( \bar{\lambda}_+ R_{1 j} R_{1 k} + \bar{\lambda}_- R_{3 j} R_{3 k} + 2 \lambda_{H \eta} R_{2 j} R_{2 k} \right) v_H + \mathcal{R}_{j k}^{(1, 2)} {\rm Re}(a)}$\\
$ $ & ${- \mathcal{R}_{j k}^{(2, 3)} {\rm Im}(a) - {\rm Im}(\lambda''_{H \phi}) \mathcal{R}_{j k}^{(1, 3)} v_H}$ \\ \hline
\end{tabular}
\caption{Scalar trilinear couplings defined in Eq.~\eqref{eq:relevantterms}. 
We define $\mathcal{R}_{j k}^{(a, b)} \equiv R_{a j} R_{b k} + R_{a k} R_{b j}$ and $j \neq k$ in the second row.}
\label{tab:Aij}
\end{table}
From these couplings, we can calculate the radiative mass and effective Yukawa coupling of the tau lepton for the RS model,
\begin{align}
\mathcal{L}_{\rm eff} &\supset - m_{\tau}^{\rm rad} \bar{\tau}_L \tau_R - \frac{y_{\tau}^{\rm eff}}{\sqrt{2}} \bar{\tau}_L \tau_R h^0 + {\rm h.c.} \, , \\[1.2ex]
m_{\tau}^{\rm rad} &= \sum_{j = 1}^3 \frac{y_L^{(j)} y_R^{(j)}}{16 \sqrt{2} \pi^2} m_{\psi} \left( \frac{m_{\psi}^2 B_0 (0, m_{\psi}^2, m_{\psi}^2) - m_{\varphi_j}^2 B_0 (0, m_{\varphi_j}^2, m_{\varphi_j}^2)}{m_{\varphi_j}^2 - m_{\psi}^2} - 1 \right) \nonumber \\[0.6ex]
&= \frac{y_{\phi} y_{\eta}}{16 \sqrt{2} \pi^2} \mathcal{F}_{\rm RS} (x_1, x_2, x_3) \, , \label{eq:radiativemtauRS} \\[1.0ex]
y_{\tau}^{\rm eff} (p_{h^0}^2) &= - \sum_{j, k} \frac{y_L^{(j)} y_R^{(k)} A_{j k}}{16 \pi^2} m_{\psi} C_0 (m_{\tau}^2, p_{h^0}^2, m_{\tau}^2, m_{\psi}^2, m_{\varphi_j}^2, m_{\varphi_k}^2) \, , \label{eq:YefftauRS}
\end{align}
where $\mathcal{F}_{\rm RS} (x_1, x_2, x_3)$ is defined by
\begin{align}
\mathcal{F}_{\rm RS} (x_1, x_2, x_3) &\equiv m_{\psi} \sum_{j = 1}^3 \left( R_{1 j} - i R_{3 j} \right) R_{2 j} I_{\rm rad} (x_j) \, ,
\label{eq:FfuncRS}
\end{align}
with $x_j \equiv m_{\varphi_j}^2 / m_{\psi}^2$. 
Note that $\mathcal{F}_{\rm RS} (x_1, x_2, x_3)$ is scaled by $m_{\psi}$, as expected in Eq.~\eqref{eq:mtauradGen}. 
By using orthogonal conditions, $\sum_{m = 1}^3 R_{jm} R_{km} = \delta_{jk} = \sum_{m = 1}^3 R_{mj} R_{mk}$, the absolute value of $\mathcal{F}_{\rm RS} (x_1, x_2, x_3)$ is given by
\begin{align}
|\mathcal{F}_{\rm RS} (x_1, x_2, x_3)| &= \sqrt{|\mathcal{F}_{\rm RS} (x_1, x_2, x_3)|^2} \nonumber \\[0.3ex]
&= m_{\psi} \Bigl( \Bigr. R_{21}^2 R_{22}^2 \left[ I_{\rm rad} (x_1) - I_{\rm rad} (x_2) \right]^2 + R_{22}^2 R_{23}^2 \left[ I_{\rm rad} (x_2) - I_{\rm rad} (x_3) \right]^2 \nonumber \\[0.3ex]
&\hspace{4.5em} + R_{23}^2 R_{21}^2 \left[ I_{\rm rad} (x_3) - I_{\rm rad} (x_1) \right]^2 \Bigl. \Bigr)^{1/2} \, . \label{eq:FcalRSsimple}
\end{align}

$C_T (0)$ and $C_{T'} (0)$ in the RS model can be calculated from the right diagram of Fig.~\ref{fig:dipolediagrams},
\begin{align}
C_T^{\rm RS} (0) &= \sum_{j = 1}^3 \frac{{\rm Re}[y_L^{(j)} y_R^{(j)}]}{16 \sqrt{2} \pi^2} m_{\psi} C_1 (m_{\tau}^2, 0, m_{\tau}^2, m_{\varphi_j}^2, m_{\psi}^2, m_{\psi}^2) \, , \label{eq:CT0RS} \\[1ex]
C_{T'}^{\rm RS} (0) &= \sum_{j = 1}^3 \frac{{\rm Im}[y_L^{(j)} y_R^{(j)}]}{16 \sqrt{2} \pi^2} m_{\psi} C_1 (m_{\tau}^2, 0, m_{\tau}^2, m_{\varphi_j}^2, m_{\psi}^2, m_{\psi}^2) \, . \label{eq:CTp0RS}
\end{align}
Note that in the limit of $m_{\tau}^2 \ll m_{\psi}^2, m_{\varphi_j}^2$, we are able to find the analytical expression of $C_1 (m_{\tau}^2, 0, m_{\tau}^2, m_{\varphi_j}^2, m_{\psi}^2, m_{\psi}^2) \approx C_1 (0, 0, 0, m_{\varphi_j}^2, m_{\psi}^2, m_{\psi}^2)$ as
\begin{align}
C_1 (0, 0, 0, m_{\varphi_j}^2, m_{\psi}^2, m_{\psi}^2) = \frac{1}{m_{\psi}^2} \frac{3 x_j^2 - 4 x_j + 1 - 2 x_j^2 \ln x_j}{4 (1 - x_j)^3} \, .
\end{align}
With Eqs.~\eqref{eq:atauGen}, \eqref{eq:dtauGen}, \eqref{eq:CT0RS} and \eqref{eq:CTp0RS}, the dipole moments $a_{\tau}$ and $d_{\tau}$ can be expressed as
\begin{align}
a_{\tau} &= 2 \left( \frac{m_{\tau}^{\rm exp}}{|\mathcal{F}_{\rm RS} (x_1, x_2, x_3)|} \right)^2 \Bigl\{ \Bigr. R_{21}^2 R_{22}^2 \bigl[ \widetilde{C}_1 (1) - \widetilde{C}_1 (2) \bigr] \bigl[ I_{\rm rad} (x_1) - I_{\rm rad} (x_2) \bigr] \nonumber \\[0.3ex]
&\hspace{11.0em} + R_{22}^2 R_{23}^2 \bigl[ \widetilde{C}_1 (2) - \widetilde{C}_1 (3) \bigr] \bigl[ I_{\rm rad} (x_2) - I_{\rm rad} (x_3) \bigr] \nonumber \\[1ex]
&\hspace{11.0em} + R_{23}^2 R_{21}^2 \bigl[ \widetilde{C}_1 (3) - \widetilde{C}_1 (1) \bigr] \bigl[ I_{\rm rad} (x_3) - I_{\rm rad} (x_1) \bigr] \Bigl. \Bigr\} \, , \label{eq:atauRSsimple} \\[1ex]
d_{\tau} &= - \frac{m_{\tau}^{\rm exp}}{|\mathcal{F}_{\rm RS} (x_1, x_2, x_3)|^2} R_{21} R_{22} R_{23} I_d^{\rm RS} (x_1, x_2, x_3) \, , \label{eq:dtauRSsimple}
\end{align}
where $\widetilde{C}_1 (j) \equiv \widetilde{C}_1 (m_{\tau}^2, m_{\varphi_j}^2, m_{\psi}^2)$ with $\widetilde{C}_1 (m_{\tau}^2, m_{\varphi_j}^2, m_{\psi}^2)$ in appendix~\ref{app:PVint}, and we define
\begin{align}
I_d^{\rm RS} (x_1, x_2, x_3) &\equiv \widetilde{C}_1 (1) \bigl[ I_{\rm rad} (x_2) - I_{\rm rad} (x_3) \bigr] + \widetilde{C}_1 (2) \bigl[ I_{\rm rad} (x_3) - I_{\rm rad} (x_1) \bigr] \nonumber \\[0.3ex]
&\hspace{5.0em} + \widetilde{C}_1 (3) \bigl[ I_{\rm rad} (x_1) - I_{\rm rad} (x_2) \bigr] \, . \label{eq:IdRSsimple}
\end{align}
Here, for $d_{\tau}$, we have also used the following identities for the orthogonal matrix, in addition to the orthogonal conditions:
\begin{align}
R_{13} R_{32} - R_{12} R_{33} = R_{21} \, , ~~ R_{11} R_{33} - R_{13} R_{31} = R_{22} \, , ~~ R_{12} R_{31} - R_{11} R_{32} = R_{23} \, .
\end{align}
As a result, $a_{\tau}$ and $d_{\tau}$ only depend on $R_{21, 22, 23}$ in addition to the exotic masses. 
Due to the orthogonal condition $R_{21}^2 + R_{22}^2 + R_{23}^2 = 1$, only two of them are independent parameters, which leads to one remaining free parameter in $R$. 
Note that we numerically check that $\widetilde{C}_1 (j) > \widetilde{C}_1 (k)$ when $m_{\varphi_j} < m_{\varphi_k}$, and $I_{\rm rad} (x)$ is the monotonically increasing function, and therefore, $\bigl[ \widetilde{C}_1 (j) - \widetilde{C}_1 (k) \bigr] \bigl[ I_{\rm rad} (x_j) - I_{\rm rad} (x_k) \bigr] < 0$ for $m_{\varphi_j} < m_{\varphi_k}$, which results in a negative value of $a_{\tau}$ in the RS model.

\section{Phenomenology}
\label{sec:pheno}

We here discuss experimental constraints on our model, which will restrict its parameter space for generating large tau dipole moments $a_{\tau}$ and $d_{\tau}$.

\subsection{Higgs decay to tau leptons}
\label{sec:h2tautau}

The current experimental result on branching ratio of $h^0 \to \tau^+ \tau^-$ reported by ATLAS in 2023 is~\cite{ATLAS:2022vkf}
\begin{align}
{\rm BR} (h^0 \to \tau^+ \tau^-) \equiv \frac{\Gamma_{h^0 \to \tau^+ \tau^-}}{\Gamma_{h^0, {\rm total}}} = 0.060^{+0.008}_{-0.007} \, ,
\label{eq:BRh2tautauExp}
\end{align}
which is quite perfectly predicted by SM, because its uncertainty is much smaller than that of $h^0 \to \mu^+ \mu^-$ channel. 
This means that our effective Yukawa coupling $y_{\tau}^{\rm eff} (m_{h^0}^2)$ should not largely deviate from the SM Yukawa coupling of the tau lepton $y^{\rm SM}_{\tau}$, which will be a strong constraint on our parameter space. 

In our model, the decay width of $h^0 \to \tau^+ \tau^-$ is estimated as
\begin{align}
\Gamma_{h^0 \to \tau^+ \tau^-} = \frac{m_{h^0}}{16 \pi} \sqrt{1 - \frac{4 m_{\tau}^2}{m_{h^0}^2}} \left[ \left( 1 - \frac{4 m_{\tau}^2}{m_{h^0}^2} \right) ({\rm Re} \, y_{\tau}^{\rm eff})^2 + ({\rm Im} \, y_{\tau}^{\rm eff})^2 \right] \, .
\label{eq:Gamh2tautau}
\end{align}
In the limit of $4 m_{\tau}^2 \ll m_{h^0}^2$, we can simply get \begin{align}
\Gamma_{h^0 \to \tau^+ \tau^-} \simeq \frac{m_{h^0}}{16 \pi} |y_{\tau}^{\rm eff}|^2 \, .
\end{align}
The total decay width of Higgs from SM prediction is given by $\Gamma_{h^0, {\rm total}}^{\rm SM} = 4.115$ MeV, when Higgs mass is $m_{h^0} = 125.2$ GeV~\cite{LHCHiggsCrossSectionWorkingGroup:2016ypw}. 
According to the ATLAS result, we find the region of $\Gamma_{h^0 \to \tau^+ \tau^-} \in (0.216, 0.281)$ MeV for $1\sigma$ interval of the experimental result in Eq.~\eqref{eq:BRh2tautauExp}.\footnote{To obtain this $1\sigma$ range, we also use the SM prediction of ${\rm BR} (h^0 \to \tau^+ \tau^-) = 0.0624$ for $m_{h^0} = 125.2$ GeV in ref.~\cite{LHCHiggsCrossSectionWorkingGroup:2016ypw}, in order to subtract the SM contribution of $h^0 \to \tau^+ \tau^-$ from $\Gamma_{h^0, {\rm total}}^{\rm SM}$.}
The constraint of $|y_{\tau}^{\rm eff}|$ is given by 
\begin{align}
|y_{\tau}^{\rm eff} (m_{h^0}^2)| \in (0.93,1.06) \times 10^{-2} \, .
\label{eq:yeffrange}
\end{align}
Clearly, the SM value $y_{\tau}^{\rm SM} = \sqrt{2} m_{\tau} / v_H \simeq 1.02 \times 10^{-2}$ also lies in this region.

\subsection{Higgs invisible decay}
\label{sec:h2inv}

Since our models have couplings between the SM Higgs and new exotic particles, some parameter space suffers from the constraint of the Higgs invisible decay, whose bound is~\cite{ParticleDataGroup:2024cfk}
\begin{align}
\frac{\Gamma_{h^0, \, {\rm inv}}}{\Gamma_{h^0, {\rm total}}} < 0.107 ~~ (95 \% \: {\rm C.L.}) \, .
\label{eq:htoinv}
\end{align}
Using the total decay width of the SM Higgs, the constraint can be $\Gamma_{h^0, \, {\rm inv}} < 0.493$ MeV, if $|y_{\tau}^{\rm eff}| \simeq y_{\tau}^{\rm SM}$. 
When using the obtained range of $|y_{\tau}^{\rm eff}|$ in Eq.~\eqref{eq:yeffrange}, the bound becomes $\Gamma_{h^0, \, {\rm inv}} < 0.488$-$0.496$ MeV. 

The MF and RS models have following couplings which are relevant to $\Gamma_{h^0, \, {\rm inv}}$: tree-level scalar trilinear couplings of $h^0 \varphi_j^- \varphi_k^+$ or $h^0 \varphi_j \varphi_k$, and one-loop couplings of $h^0 \overline{\psi}_a \psi_b$ or $h^0 \overline{\psi} \psi$. 
The lightest scalar $\varphi_1$ or fermion $\psi_1$ ($\psi$ for the RS model) can be lighter than the SM Higgs, and hence, these couplings lead to decay processes of $h^0 \to \varphi_1 \varphi_1$ or $h^0 \to \psi_1 \psi_1$. 
The decay width for the former process is given by
\begin{align}
\Gamma_{h^0, \, {\rm inv}} = \Gamma_{h^0 \to \varphi_1 \varphi_1} = \frac{|A_{11}|^2}{32 \pi m_{h^0}} \sqrt{1 - \frac{4 m_{\varphi_1}^2}{m_{h^0}^2}} \, ,
\label{eq:GamhtoinvS}
\end{align}
where $A_{11}$ depends on the model and can be found in Tables~\ref{tab:yLRAjk} and \ref{tab:Aij} for the MF model and RS model, respectively. 
On the other hand, the decay width for the latter process can be found in a similar manner as in Eq.~\eqref{eq:Gamh2tautau}:
\begin{align}
\Gamma_{h^0, \, {\rm inv}} = \Gamma_{h^0 \to \psi_1 \psi_1} = \frac{m_{h^0}}{16 \pi} C_{h^0, \psi_1} \sqrt{1 - \frac{4 m_{\psi_1}^2}{m_{h^0}^2}} \left[ \left( 1 - \frac{4 m_{\psi_1}^2}{m_{h^0}^2} \right) ({\rm Re} \, y^{\rm eff}_{\psi_1})^2 + ({\rm Im} \, y^{\rm eff}_{\psi_1})^2 \right] \, ,
\label{eq:Gamh2psipsi}
\end{align}
where $C_{h^0, \psi_1} = 1/2 \, (1)$ for the MF (RS) model is the symmetry factor, $\psi_1 = \psi$ for the RS model, and $y_{\psi_1}^{\rm eff}$ can be estimated by replacing $m_{\tau} \leftrightarrow m_{\psi_1}$ and $y_L \leftrightarrow y_R$ in Eqs.~\eqref{eq:YefftauMF} and \eqref{eq:YefftauRS}. 
Note that due to this replacement, $y_{\psi_1}^{\rm eff}$ will be small, because it is proportional to $m_{\tau}$, instead of $m_{\psi}$. 
It is emphasized that only when $m_{\varphi_1} < m_{h^0} / 2$ or $m_{\psi_1} < m_{h^0} / 2$, these invisible decays are kinematically allowed.

\subsection{Collider bounds on charged particles}
\label{sec:massbounds}

In both models, we introduce new $U(1)_{\rm em}$ charged particle(s), namely, charged scalars in the MF model or a charged fermion in the RS model. 
The masses of new charged particles are constrained by collider searches. 
The LEP experiment puts a bound on the charged scalar mass,\footnote{If the new charged scalar couples to the quark sector, other bounds should be considered. 
See, e.g., refs.~\cite{ATLAS:2014otc,CMS:2015lsf} for $m_{H^+} < m_t$ and ref.~\cite{ATLAS:2018gfm} for $m_{H^+} > m_t$.} $m_{H^+} > 94$ GeV at 95\% C.L.~\cite{ALEPH:2013htx}, for the case of BR$(H^+ \to \tau^+ \nu) = 1$ in type-II two Higgs doublet model (2HDM). 
They also concluded that $m_{H^+} < 80$ GeV is excluded for the type-II 2HDM, for any value of BR$(H^+ \to \tau^+ \nu)$. 
These bounds will be relevant to the MF model. 

A new charged fermion is also constrained by the LEP experiment. 
The lower bound on a new charged fermion mass is $100.5$-$102.6$ GeV, depending on the charge assignment of the new fermion~\cite{L3:2001xsz}. 
In addition, the CMS experiment gives a bound on the vector-like lepton mass~\cite{CMS:2022nty}. 
The singlet vector-like lepton case in their analyses is relevant for the RS model, and gives the exclusion range of $125 \, {\rm GeV} < m_{\tau'} < 150$ GeV. 
Here, $\tau'$ is a new singlet vector-like lepton, which corresponds to $\psi$ in the RS model. 

We can also consider constraints from searches for stau production at the LHC~\cite{ATLAS:2014hep,ATLAS:2015eiz,ATLAS:2019gti,CMS:2016gjw,CMS:2019eln,CMS:2022syk}. 
The direct stau production constrains a region for the stau mass below 480 GeV, depending on the neutralino mass up to $\simeq 215$ GeV~\cite{ATLAS:2023djh}. 
This constraint should be applied to the parameter space of the MF model, since there are stau-like ($\varphi_{1, 2}^{\pm}$) and neutralino-like ($\psi_{1, 2}$) exotic particles in the model.

\subsection{Lepton-flavor universality of $Z$ boson decays}
\label{sec:LFUZdecays}

Since our exotic particles exclusively couple to the tau lepton, the decay width of $Z \to \tau^+ \tau^-$ is modified in our model. 
Due to this, the ratio between the $Z \to e^+ e^-, \mu^+ \mu^-$ and $Z \to \tau^+ \tau^-$ decay widths may restrict our parameter space. 
The current experimental status for these ratios is~\cite{ALEPH:2005ab,LHCb:2018ogb}\footnote{Here, we show the averaged results in ref.~\cite{ParticleDataGroup:2024cfk} and use them for numerical analyses.}
\begin{align}
\frac{\Gamma(Z \to \tau^+ \tau^-)}{\Gamma(Z \to e^+ e^-)} &= 1.0020 \pm 0.0032 \, , \label{eq:Gamtau/Game} \\[1ex]
\frac{\Gamma(Z \to \tau^+ \tau^-)}{\Gamma(Z \to \mu^+ \mu^-)} &= 1.0010 \pm 0.0026 \, . \label{eq:Gamtau/Gammu}
\end{align}
We call these bounds lepton-flavor universality (LFU) constraints. 

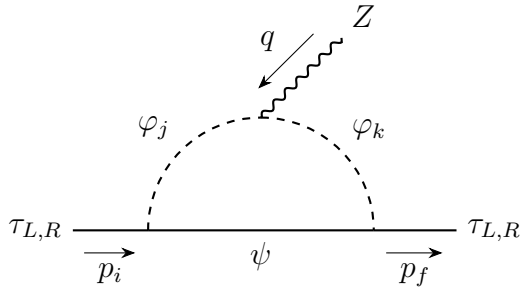
\begin{figure}[!t]
\begin{center}
\begin{tikzpicture}
\begin{feynman}[large]
\vertex (a1) {\(\tau_{L, R}\)};
\vertex [right=1.5cm of a1] (b1);
\vertex [right=3cm of b1] (c1);
\vertex [right=1.1cm of c1] (d1) {\(\tau_{L, R}\)};
\vertex [above right=2.12132cm of b1] (f1);
\vertex [above right=1.5cm of f1] (g1) {\(Z\)};
\diagram [medium] {
(a1) -- [momentum'=\(p_i\)] (b1) -- [edge label'=\(\psi\)] (c1) -- [momentum'=\(p_f\)] (d1),
(b1) -- [scalar, quarter left, edge label=\(\varphi_j\)] (f1) -- [scalar, quarter left, edge label=\(\varphi_k\)] (c1),
(f1) -- [photon, rmomentum=\(q\)] (g1),
};
\end{feynman}
\end{tikzpicture}
\end{center}
\vspace{-0.5cm}
\caption{Additional diagram related to $Z \to \tau^+ \tau^-$ decay process for the RS model. 
Due to the $Z$ boson coupling structure, only $j \neq k$ cases contribute to the decay process. }
\label{fig:dipolediagramZ}
\end{figure}

The tau lepton couplings to the $Z$ boson can be parameterized as 
\begin{align}
\mathcal{L}_Z \supset \frac{g}{c_w} \bar{\tau} \gamma^{\alpha} \Bigl[ (g_L^{\tau} + \delta g_L^{\tau}) P_L + (g_R^{\tau} + \delta g_R^{\tau}) P_R \Bigr] \tau Z_\alpha \, , \label{eq:LZtautau}
\end{align}
where $g$ denotes the $SU(2)_L$ gauge coupling, $c_w$ is the cosine of the weak mixing angle $\theta_W$, and $g_L^{\tau} = - \frac{1}{2} + s_w^2$, $g_R^{\tau} = s_w^2$ with $s_w \equiv \sin \theta_W$ are the tree-level tau couplings to the $Z$ boson in the SM. 
In our models, new physics contributions $\delta g_{L,R}^{\tau}$ are induced by each dipole diagram replacing the photon to the $Z$ boson. 
For the RS model, $Z$-$\varphi_j$-$\varphi_k$ ($j \neq k$) couplings exist, and hence, an additional diagram shown in Fig.~\ref{fig:dipolediagramZ} should be considered. 
The ratios can be then estimated as
\begin{align}
\frac{\Gamma(Z \to \tau^+ \tau^-)}{\Gamma(Z \to \ell^+ \ell^-)} \simeq 1 + \frac{2 g_L^{\ell} {\rm Re} (\delta g_L^\tau) + 2 g_R^{\ell} {\rm Re} (\delta g_R^\tau)}{(g_L^{\ell})^2+(g_R^{\ell})^2}\equiv 1+ \delta_{\tau}^{\ell} \, ,
\end{align}
where $g_{L,R}^{\ell} = g_{L,R}^{\tau}$ are the universal lepton couplings to the $Z$ boson in the SM, and we assume the new physics contributions are smaller than those of the SM, $\delta g_{L,R}^{\tau} \ll g_{L,R}^{\tau}$. 
Therefore, $|\delta_{\tau}^{\ell}|$ must be less than $\mathcal{O}(10^{-3})$: $-0.0012 < \delta_{\tau}^{\ell} < 0.0036$ ($-0.0042 < \delta_{\tau}^{\ell} < 0.0062$) is required for within $1\sigma$ ($2\sigma$) range to satisfy both constraints in Eqs.~\eqref{eq:Gamtau/Game} and \eqref{eq:Gamtau/Gammu}. 
The full expressions for $\delta g^\tau_{L, R}$ are summarized in appendix~\ref{app:Ztautau-full}.

\subsection{Electroweak precision measurement}
\label{sec:EWPO}

Since our exotic particles are coupled to the SM gauge bosons, there are additional loop corrections to the propagators of the gauge bosons. 
This leads to non-zero contributions to oblique parameters $S, T, U$ and finally results in a possible deviation of the $W$ boson mass from the SM prediction. 
The current bounds on these quantities can be found in ref.~\cite{STU:2016ojx} as
\begin{align}
S &\simeq 0.09 \pm 0.10 \, , \\
T &\simeq 0.10 \pm 0.12 \, , \\
U &\simeq 0.01 \pm 0.09 \, , \\
m_W &\simeq 80.3692 \pm 0.0133 \, {\rm GeV} \, ,
\end{align}
where the $W$ boson mass is taken from ref.~\cite{ParticleDataGroup:2024cfk}. 
In the MF model, only exotic scalars contribute to these quantities, because exotic fermions $\psi_a$ are totally singlet under the SM gauge groups. 
On the other hand, in the RS model, all exotic particles have contributions to them. 
In appendix~\ref{app:STU}, we summarize details of contributions to $S, T, U$ parameters for each model.

\section{Results}
\label{sec:results}

In this section, we show the numerical results in both the MF model and RS model. 
In our analyses, we focus on the on-shell photon case, $q^2 = 0$, although collider experiments can extract the information of the form factor with the off-shell photon. 
We will comment on this in section~\ref{sec:summary}.

\subsection{MF model}
\label{sec:MFresult}

For the numerical analyses of the MF model, we choose the relevant parameters as
\begin{align}
&M_{\phi}^2 = (200 \, {\rm GeV})^2 \, , ~~ a = 120 \, {\rm GeV} \, , ~~ m_D = 200 \, {\rm GeV} \, , ~~ m_{RR} = 200 \, {\rm GeV} \, , ~~ \theta_{\rm phys}^{\rm MF} = 0.8 \, , \nonumber \\
&(150 \, {\rm GeV})^2 \leq M_{\eta}^2 \leq (450 \, {\rm GeV})^2 \, , ~~ 5 \, {\rm GeV} \leq m_{LL} \leq 150 \, {\rm GeV} \, . \label{eq:ParamsMF}
\end{align}
Note that for this parameter choice, relevant constraints on the parameter space will come from: (i) the LEP bound on the charged scalar, (ii) the constraint on the combination $y_{\phi} y_{\eta} < 4 \pi$. 
The other constraints can be imposed by setting other parameters, $y_{\phi}$ (or $y_{\eta}$), $\lambda_{H \phi}$, $\lambda_{H \eta}$, $\lambda'_{H \phi}$ and $\lambda''_{H \phi}$, which do not affect the results of $a_{\tau}$ and $|d_{\tau}|$. 

\begin{figure}[!t]
\centering
\includegraphics[width=0.48\textwidth]{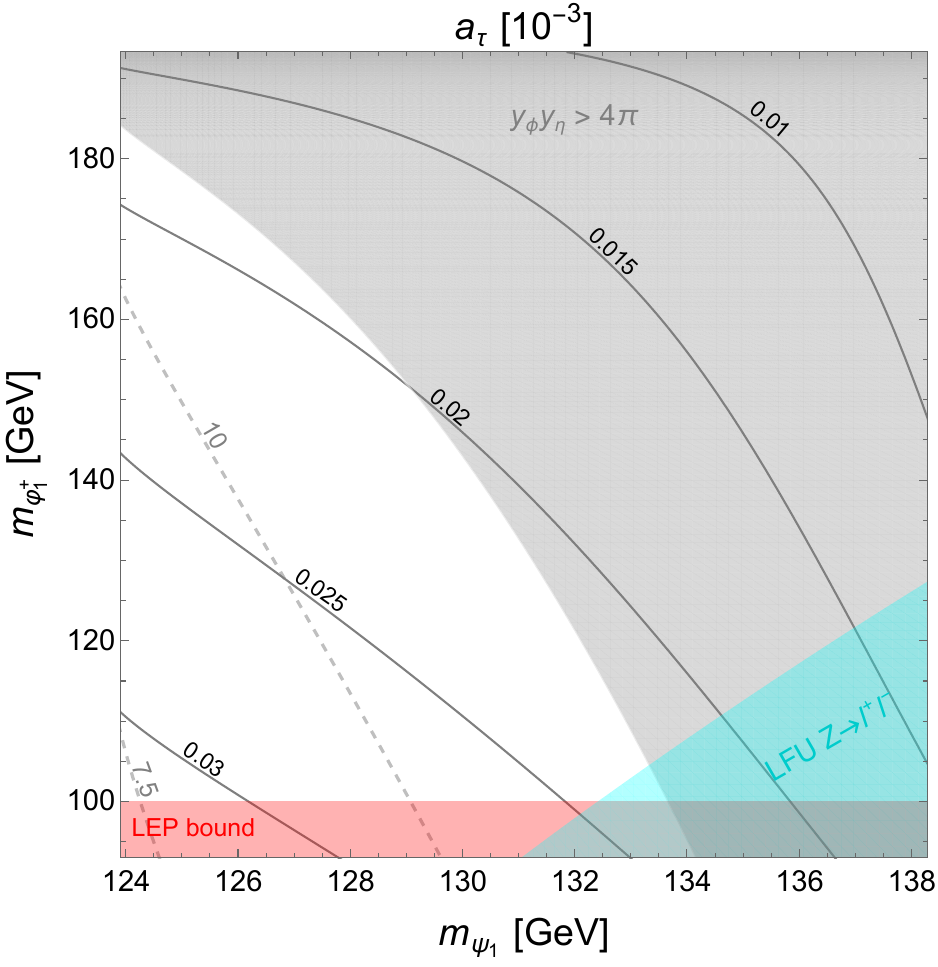} ~~ \includegraphics[width=0.48\textwidth]{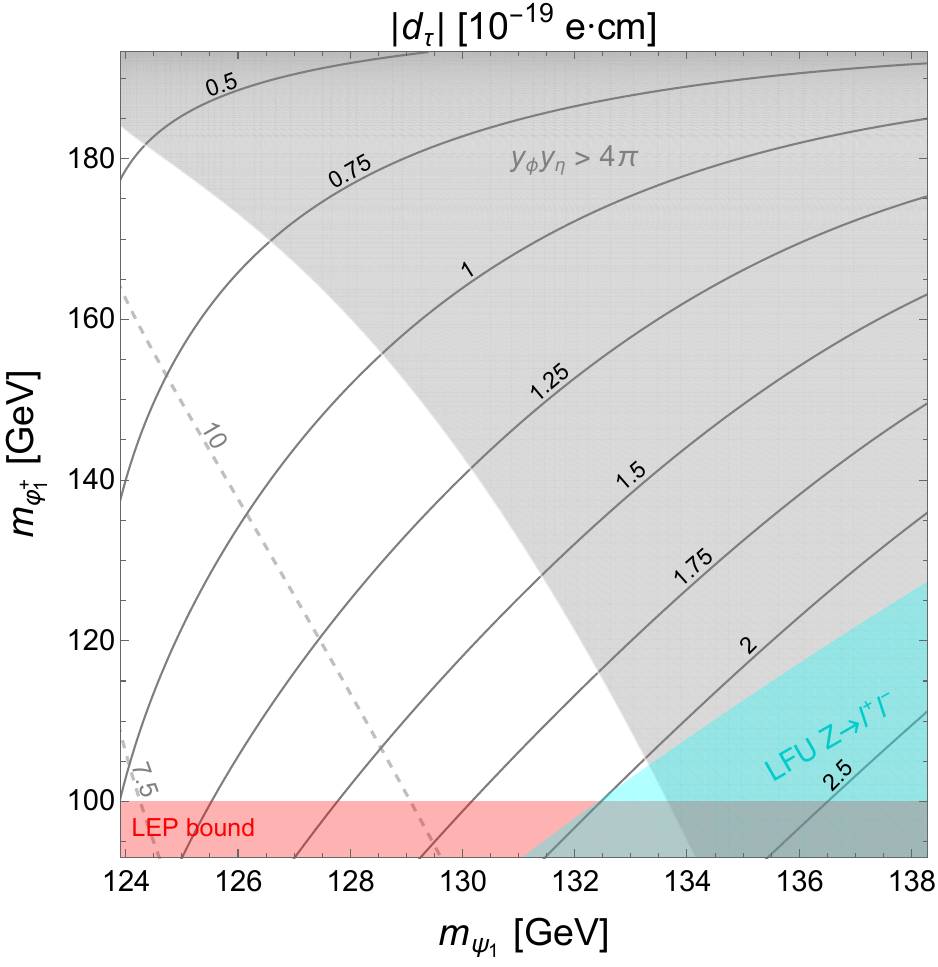}
\caption{Numerical predictions of $a_{\tau}$ (left panel) and $|d_{\tau}|$ (right panel) in the MF model with Eq.~\eqref{eq:ParamsMF} and $(150 \, {\rm GeV})^2 \leq M_{\eta}^2 \leq (450 \, {\rm GeV})^2$ and $5 \, {\rm GeV} \leq m_{LL} \leq 150 \, {\rm GeV}$. 
Each black contour shows the predictions of $a_{\tau}$ and $|d_{\tau}|$, while the gray dashed line is the contour for the combination of Yukawa couplings, $y_{\phi} y_{\eta}$. 
The shaded regions are excluded by the LEP bound (red), $y_{\phi} y_{\eta} > 4 \pi$ (gray) and the LFU constraint (cyan) by setting $y_{\phi} = y_{\eta}$. }
\label{fig:tauMDMandEDM}
\end{figure}

Fig.~\ref{fig:tauMDMandEDM} shows the numerical results in the MF model with parameters in Eq.~\eqref{eq:ParamsMF}, in $(m_{\psi_1}, m_{\varphi_1})$-plane. 
The left panel is predictions of $a_{\tau}$, while the right one is $|d_{\tau}|$. 
The red shaded region is the LEP bound on the charged scalar mass, and we take a conservative limit, $m_{\varphi_1^+} > 100$ GeV. 
The gray dashed lines are the contours for the combination of Yukawa couplings, $y_{\phi} y_{\eta}$, and the gray shaded region corresponds to $y_{\phi} y_{\eta} > 4 \pi$. 
We also plot the LFU constraint on the model by setting $y_{\phi} = y_{\eta}$, as shown in the cyan shaded region. 

From the left panel of Fig.~\ref{fig:tauMDMandEDM}, $a_{\tau} \simeq (2 \mathchar`- 3) \times 10^{-5}$ is predicted in the whole allowed parameter space, and therefore, our model with the parameter choice in Eq.~\eqref{eq:ParamsMF} will be tested in future experiments. 
The right panel of Fig.~\ref{fig:tauMDMandEDM} shows that a large $|d_{\tau}| > 10^{-19} \ecm$ can be obtained for $m_{\psi_1} \simeq 124$-$133$ GeV with $100 \, {\rm GeV} \lesssim m_{\varphi_1^+} \lesssim 155 \, {\rm GeV}$, and future measurements of $|d_{\tau}|$ will be able to probe this region. 
Importantly, the value of $|d_{\tau}|$ is close to the indirect bound in Eq.~\eqref{eq:tauEDMindirect}, and the region of $|d_{\tau}| > 10^{-19} \ecm$ can be also tested by electron EDM measurements. 
Note that this result is consistent with ref.~\cite{Khaw:2022qxh} for the muon EDM, which gives $|d_{\mu}| \simeq 10^{-22} \ecm$ for $m_{\psi_1}, m_{\varphi_1^+} \simeq 720$ GeV: the expected value of $|d_{\tau}|$ is then $|d_{\tau}| \sim |d_{\mu}| \times (m_{\tau} / m_{\mu}) \times (720 \, {\rm GeV} / 130 \, {\rm GeV} )^2 \sim 2 \times 10^{-19} \ecm$. 

It is clear that the constraint of $y_{\phi} y_{\eta} < 4 \pi$ is crucial for the viable parameter space in the MF model. 
As mentioned in section~\ref{sec:setup}, this is because we reproduce the tau lepton mass in the current work, which is about 17 times heavier than the muon mass. 
Moreover, we focus on the lighter mass region for the exotic particles, $\mathcal{O} (100)$ GeV, and hence, $y_{\phi} y_{\eta}$ is roughly two orders of magnitude larger than that in the radiative muon mass model~\cite{Khaw:2022qxh}, which will be also expected in the RS model. 
For a heavier $m_{\varphi_1^+}$ region which is obtained for a larger $M_{\eta}^2$ with fixed $M_{\phi}^2$ and $a$, the value of $y_{\phi} y_{\eta}$ increases, because the mixing angle $\theta$ decreases. 
This can be understood from the definition of each mass eigenvalue in Eqs.~\eqref{eq:eigenpm1} and \eqref{eq:eigenpm2}: when $M_{\eta}^2 \gg M_{\phi}^2, a v_H$, each mass eigenvalue is $m_{\varphi_1^+}^2 \simeq M_{\phi}^2$ and $m_{\varphi_2^+}^2 \simeq M_{\eta}^2$, and therefore, $\sin 2 \theta$ decreases by $\sim 1 / M_{\eta}^2$. 
The $m_{\psi_1}$ dependence of $y_{\phi} y_{\eta}$ is, on the other hand, not so simple, although $\sin 2 \alpha$ decreases when $m_{LL} \gg m_{RR}, m_D$. 
In the plot region of Fig.~\ref{fig:tauMDMandEDM}, however, $s_{2 \alpha} \simeq 0.89$-$0.98$ due to the parameter choice in Eq.~\eqref{eq:ParamsMF}. 
Therefore, $y_{\phi} y_{\eta}$ mainly depends on the size of $|\mathcal{F}_{\rm MF} (x_{1, 1}, x_{1, 2}, x_{2, 1}, x_{2, 2})|$, which can be estimated by
\begin{align}
|\mathcal{F}_{\rm MF} (x_{1, 1}, x_{1, 2}, x_{2, 1}, x_{2, 2})| &= \sqrt{m_{\psi_1}^2 I_1^2 + m_{\psi_2}^2 I_2^2 - 2 m_{\psi_1} m_{\psi_2} I_1 I_2 \cos 2 \theta_{\rm MF}} \, , \\[0.5ex]
\text{with }~ I_a &\equiv I_{\rm rad} (x_{1, a}) - I_{\rm rad} (x_{2, a}) < 0 \, .
\end{align}
Here, $\theta_{\rm MF}$ is determined by Eq.~\eqref{eq:tantau}, and hence, it decreases when $m_{LL} = 5 \, {\rm GeV} \to 150$ GeV with $m_{\rm RR} = 200$ GeV, leading to $\cos 2 \theta_{\rm MF} \to 1$. 
Then, $|\mathcal{F}_{\rm MF} (x_{1, 1}, x_{1, 2}, x_{2, 1}, x_{2, 2})|$ decreases when $m_{LL}$ increases within the range of $m_{LL} < m_{RR}$. 
This results in a large $y_{\phi} y_{\eta}$ for the region of heavy $m_{\psi_1}$. 

We have comments on the other constraints on the plots. 
The LFU constraint depends on the choice of a ratio $r_{y} \equiv y_{\phi} / y_{\eta}$, which is $r_y = 1$ in the plots. 
We checked that the $r_{y} \simeq 1$ is required for a large viable parameter space, because the dominant parts of $\delta g_{L, R}^{\tau}$ are proportional to $y_{\phi}^2, y_{\eta}^2$, respectively, as explained in appendix~\ref{sec:MFZtautau}. 
It is emphasized that the value of ratio $r_{y} \simeq 1$ is also expected from the point of view of large $y_{\phi} y_{\eta}$: in most of the viable parameter space, $y_{\phi} y_{\eta} = \mathcal{O} (10)$, and individual couplings $y_{\phi}, y_{\eta}$ are favored to be less than $\sqrt{4 \pi}$. 
For other constraints, e.g., the effective tau Yukawa coupling $y_{\tau}^{\rm eff}$ and the oblique parameters, can be satisfied by choosing quartic couplings, $\lambda_{H \phi}$, $\lambda_{H \eta}$, $\lambda'_{H \phi}$ and $\lambda''_{H \phi}$. 
We checked that $(\lambda_{H \phi}, \lambda_{H \eta}, \lambda'_{H \phi}, \lambda''_{H \phi}) = (0.15, 0.18, 0.2, 0.3)$ gives $|y_{\tau}^{\rm eff}| \simeq 0.00966$-$0.0104$, $S \simeq 0.0043$-$0.0133$, $T \simeq -0.0032$-$0.0023$, $U \simeq -0.00022$-$0.00056$, which are within the experimental bounds. 
Moreover, since the lightest exotic particle is heavier than $m_{h^0} / 2$, there is no constraint from the Higgs invisible decay.

\subsection{RS model}
\label{sec:RSresult}

Among 14 new parameters in the Lagrangian and scalar potential,
the following 11 parameters are independent and relevant to our analysis of the RS model:
\begin{align}
y_{\phi} \, , ~ y_{\eta} \, , ~ m_{\psi} \, , ~ m_{\phi}^2 \, , ~ m_{\eta}^2 \, , ~ \lambda_{H \phi} \, , ~ \lambda'_{H \phi} \, , ~ |\lambda''_{H \phi}| \, , ~ \lambda_{H \eta} \, , ~ |a| \, , ~ \theta_{\rm phys}^{\rm RS} \, ,
\end{align}
where $\theta_{\rm phys}^{\rm RS}$ is defined in Eq.~\eqref{eq:CPphysRS}. 
For the scalar sector, we can choose mass eigenvalues and mixing matrix elements $R_{jk}$ as our input parameters for numerical analysis, instead of using the parameters in the Lagrangian. 
According to the expressions for $a_{\tau}$ and $d_{\tau}$ in Eqs.~\eqref{eq:atauRSsimple} and \eqref{eq:dtauRSsimple}, we use $m_{\psi}, m_{\varphi_1}, m_{\varphi_2}, m_{\varphi_3}, R_{21}, R_{22}$ as inputs for numerical predictions of dipole moments. 
To investigate a viable parameter space, we need to fix remaining parameters, $y_{\phi}$ (or $y_{\eta}$), one of $R_{jk}$ and quartic couplings. 
For the current analysis, we choose $R_{31}$ as an input parameter, and each input is chosen as
\begin{align}
&y_{\phi} = 2 \, , ~~ 
\lambda_{H \phi} = 0.7 \, , ~~ \lambda'_{H \phi} = 0 \, , ~~ \lambda_{H \eta} = -0.15 \, , \nonumber \\[0.5ex]
&m_{\varphi_2} = 200 \, {\rm GeV} \, , ~~ m_{\varphi_3}=300 \, {\rm GeV} \, , ~~ R_{21} = 0.3 \, , ~~ R_{22} = 0.4 \, , ~~ R_{31} = 0.6 \, , \label{eq:RSpara} \\[0.5ex]
&70 \, {\rm GeV} \le m_{\varphi_1} \le 180 \, {\rm GeV} \, , ~~ 85 \, {\rm GeV} \le m_{\psi} \le 250 \, {\rm GeV} \, , \nonumber
\end{align}
and $m_{\phi}^2$, $m_{\eta}^2$, $|\lambda''_{H \phi}|$, $|a|$, and $\theta_{\rm phys}^{\rm RS}$ are reproduced by this parameter set. 
Note that $y_{\eta}$ is determined by the radiative tau mass in Eq.~\eqref{eq:radiativemtauRS}. 

\begin{figure}[!t]
\centering
\includegraphics[width=0.48\textwidth]{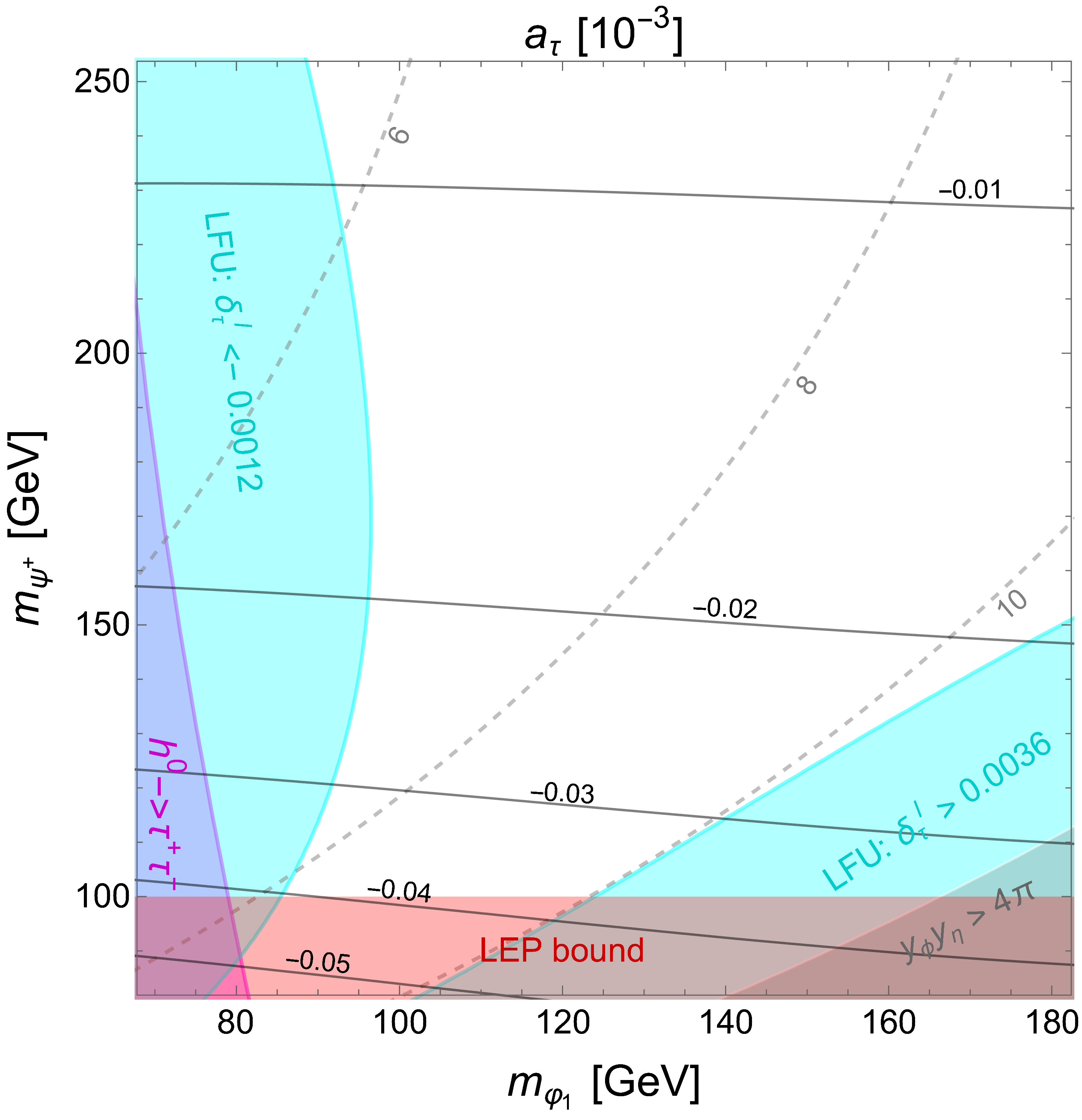} ~~ \includegraphics[width=0.48\textwidth]{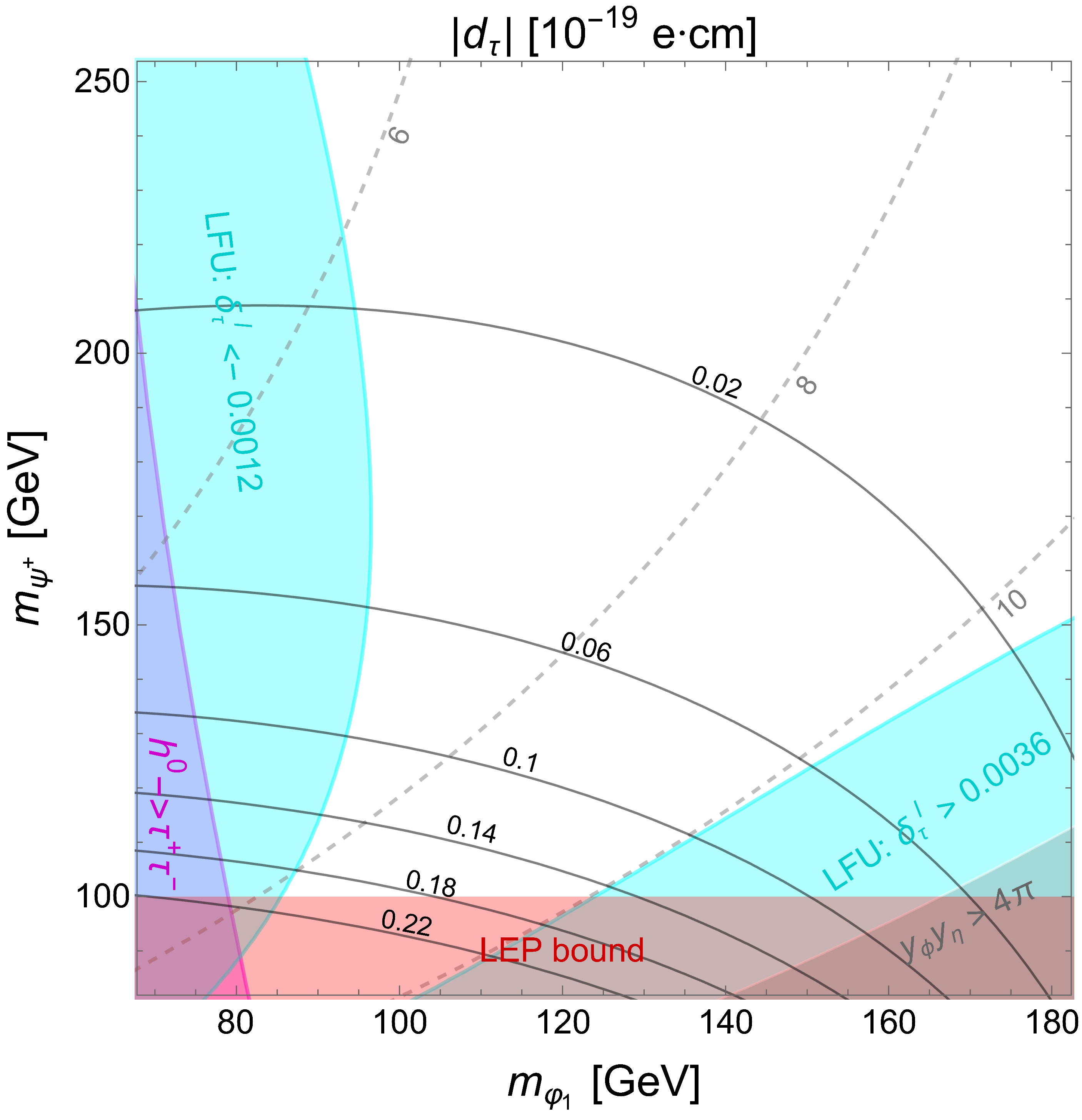}
\caption{Numerical predictions of $a_{\tau}$ (left panel) and $|d_{\tau}|$ (right panel) in the RS model with Eq.~\eqref{eq:RSpara}. 
Each black contour shows the predictions of $a_{\tau}$ and $|d_{\tau}|$, while the gray dashed lines are the contours for the combination of Yukawa couplings, $y_{\phi} y_{\eta}$. 
The shaded regions are excluded by the $h^0 \to \tau^+ \tau^-$ bound (magenta), the LEP bound (red), $y_\phi y_\eta>4\pi$ (gray) and the LFU constraint (cyan). }
\label{fig:tauMDMandEDMRS}
\end{figure}

Numerical predictions for the tau $g-2$ and EDM are shown in Fig.~\ref{fig:tauMDMandEDMRS}. 
The horizontal axis is the lightest neutral scalar mass $m_{\varphi_1}$, and the vertical one is the mass of $m_{\psi}$. 
Black solid lines correspond to contours for $a_{\tau}$ (left panel) and $d_{\tau}$ (right panel) in unit of $10^{-3}$ and $10^{-19} \ecm$, respectively. 
The shaded regions are excluded by several experimental bounds which are coming from $h^0 \to \tau^+ \tau^-$ (magenta), the LEP bound (red) and the LFU constraint (cyan). 

Compared with the MF model, there are several differences for the numerical results. 
First, $|a_{\tau}| = \mathcal{O} (10^{-5})$ is predicted in the RS model with negative value, which is positive in the MF model. 
This difference is coming from the sign of $U(1)_{\rm em}$ charge of the charged exotic particle in the loop diagram. 
This model-specific prediction is useful to distinguish the radiative mass models. 
Second, the constraint on $|y_{\tau}^{\rm eff}|$ from $h \to \tau^+ \tau^-$ appears, mainly because we are exploring a larger region compared with the MF model. 
It is important to note that we can remove it from the current plot by tuning $\lambda_{H \eta}$, without changing predictions of $a_{\tau}$ and $|d_{\tau}|$ as well as other constraints. 
We leave it here to show the bound can exclude smaller $m_{\varphi_1}$, although it is weaker than the LFU constraint. 
Third, the predictions of $a_{\tau}$ and $|d_{\tau}|$ are almost fixed by $m_{\psi}$, except for $m_{\varphi_1} \simeq 140$-$180$ GeV region of $d_{\tau}$. 
This mass dependence is expected from Eqs.~\eqref{eq:atauRSsimple} and \eqref{eq:dtauRSsimple} and the fact that $|\mathcal{F}_{\rm RS} (x_1, x_2, x_3)| \propto m_{\psi}$, which is clearly shown in the left panel of Fig.~\ref{fig:tauMDMandEDMRS}. 
For $m_{\varphi_1} \ge 140$ GeV region, the mass degeneracy between $m_{\varphi_1}$ and $m_{\varphi_2}$ gives suppression for $|d_{\tau}|$, originated from $I_d^{\rm RS} (x_1, x_2, x_3)$ function in Eq.~\eqref{eq:IdRSsimple}: $m_{\varphi_1} \simeq m_{\varphi_2}$ gives $I_d^{\rm RS} (x_1, x_2, x_3) \simeq 0$. 
As a result, $m_{\psi} \simeq m_{\varphi_1} \simeq 100$ GeV leads to large $|d_{\tau}|$ for the RS model, although it is smaller than the prediction in the MF model by one order of magnitude. 
Lastly, the constraint of $y_{\phi} y_{\eta} < 4 \pi$ is milder than the MF model and is weaker than that of the LFU constraint. 
This is because $|\mathcal{F}_{\rm RS} (x_1, x_2, x_3)|$ does not have much enhancement for $y_{\phi} y_{\eta}$ with the input values in Eq.~\eqref{eq:RSpara}. 
Note that the mass dependence of $y_{\phi} y_{\eta}$ can be understood from its analytical form in Eq.~\eqref{eq:radiativemtauRS}: smaller $m_{\psi}$ leads to larger $y_{\phi} y_{\eta}$, with fixed $m_{\varphi_{2, 3}}$ and $R_{2j}$. 
We have also checked that $|\mathcal{F}_{\rm RS} (x_1, x_2, x_3)|$ is a decreasing function with respect to $m_{\varphi_1}$, and therefore, $y_{\phi} y_{\eta} \propto 1 / |\mathcal{F}_{\rm RS} (x_1, x_2, x_3)|$ increases when $m_{\varphi_1}$ becomes heavy. 

Finally, we comment on the other constraints on the plot. 
The oblique parameter and $W$ boson mass constraints are still less important, even though there are additional fermion contributions which are absent in the MF model. 
We checked that the current parameter region gives $S \simeq 0.0021$-$0.0150$, $T \simeq -0.0168$-$0.0073$, $U \simeq -0.00099$-$0.0052$, which are within the experimental bounds. 
In addition, we should check whether the value of $|\lambda''_{H \phi}|$ is allowed by the theoretical bound, namely, the perturbative unitarity bound. 
From Eq.~\eqref{eq:massmatrix} and discussions in section~\ref{sec:RSmodel}, $|\lambda''_{H \phi}|$ can be expressed only by $m_{\varphi_{1, 2, 3}}$ and $R_{2j}$ as
\begin{align}
|\lambda''_{H \phi}| &= \sqrt{\left( 1 - R_{22}^2 \right)^2 \left( \frac{\delta m_{12}^2}{v_H^2} \right)^2 + \left( 1 - R_{23}^2 \right)^2 \left( \frac{\delta m_{13}^2}{v_H^2} \right)^2 + 2 \left( R_{22}^2 R_{23}^2 - R_{21}^2 \right) \frac{\delta m_{12}^2}{v_H^2} \frac{\delta m_{13}^2}{v_H^2}} \, ,
\label{eq:AbslamppRS}
\end{align}
where we define $\delta m_{1j}^2 \equiv m_{\varphi_j}^2 - m_{\varphi_1}^2$ ($j = 2, 3$). 
In the whole plot region of Fig.~\ref{fig:tauMDMandEDMRS}, we obtained $|\lambda''_{H \phi}| \simeq 0.273$-$0.639$, which will be safe for the bound. 
Furthermore, all exotic particles are heavier than $m_{h^0} / 2$ in the parameter space of Fig.~\ref{fig:tauMDMandEDMRS}, and hence, the Higgs invisible decay constraint is trivially satisfied.

\section{Conclusions and discussions}
\label{sec:summary}

We have explored two scenarios within a class of radiatively generated tau mass models, where new exotic fermions ($\psi$) and scalars ($\phi, \eta$) couple to the tau lepton and the SM Higgs. 
A physical CP-violating phase, unremovable by field redefinitions, naturally arises from these interactions and induces significant contributions to the tau EDM as well as the tau $g-2$. 
The models are categorized according to the hypercharge assignments of the new fermions, leading to distinct phenomenological implications. 

In the MF model obtained by setting the hypercharge $Y_{\psi} = 0$, the relevant exotic particles are charged scalars and neutral fermions. 
This setup can predict large tau EDM and $g-2$, reaching up to $a_{\tau} \sim \mathcal{O}(10^{-5})$ and $d_{\tau} \sim \mathcal{O}(10^{-19}) \ecm$, providing clear targets for upcoming experiments like the Belle II. 
These predictions lie just below the indirect limit $|d_{\tau}| < 4.1 \times 10^{-19} \ecm$ derived from electron EDM constraints via the three-loop light-by-light mechanism. 
For this reason, the MF model will be also tested by future updates of electron EDM searches. 
In contrast, the RS model obtained by the hypercharge $Y_{\psi} = -1$, which has a charged fermion and neutral scalars as relevant exotic particles, can predict the tau $g-2$ with the same order but the tau EDM one order of magnitude smaller than that of the MF model due to the suppression from the degeneracy of the exotic scalar masses. 
Another important signal is the sign of the tau $g-2$, as \textit{a model discriminator}: $a_{\tau} > 0$ favors the MF model, while $a_{\tau} < 0$ for the RS model. 
Our study demonstrates that radiative mass generation models for the tau lepton can produce testable signals in future experiments, and the precise measurements of the tau EDM and $g-2$ offer a promising avenue for uncovering new sources of CP violation and probing physics beyond the SM. 

In both models, we have a dark matter (DM) candidate which is stabilized by the unbroken $X$ symmetry: the lightest Majorana fermion $\psi_1$ in the MF model or the lightest scalar $\varphi_1$ in the RS model. 
Unless the symmetry is broken, the relic abundance should be equal or smaller than the observed DM abundance. 
If we apply the numerical result in ref.~\cite{Khaw:2022qxh}, the observed abundance may be explained via a coannihilation process in the lighter exotic mass region, $m_{\psi_1}, m_{\varphi_1} \simeq \mathcal{O} (100)$ GeV, although such light DM mass can be excluded by experiments, e.g., collider searches and direct and indirect detection bounds. 
Detailed DM phenomenology will be studied in a future work. 

Before closing, we have comments on the dipole moments of the tau lepton in the experiments. 
In our analyses, we only focus on the \textit{on-shell} photon, while the experiments can extract the information of the tau dipole moments via $e^+ e^- \to \tau^+ \tau^-$ process with momentum transfer $q^2 = E_{\rm CoM}^2$ for the \textit{off-shell} photon. 
There are some studies for optimizing the analysis methods for this process~\cite{Bernreuther:2021elu,He:2025ewk,Huang:2025ghw} and for using transverse-momentum-dependent factorization formalism to probe the dipole form factors via quasi-real photon at future lepton colliders~\cite{Shao:2025xwp}, like Super Tau-Charm Facility
(STCF)~\cite{Achasov:2023gey,Sun:2024vcd}, or at ultraperipheral heavy ion collisions~\cite{Shao:2023bga}. There are also studies discussing the light new physics contribution to $\tau$ dipole moments, through the interpretation of asymmetry measurements in $e^+ e^- \to \tau^+ \tau^-$~\cite{Hofer:2025ijh,Hofer:2025zjp}. In addition to the off-shell photon mediation, $e^+ e^- \to \tau^+ \tau^-$ is also realized by the $Z$ boson mediation, which is related to the weak dipole moments (WDMs). 
These are measured by the LEP experiment~\cite{ALEPH:2002kbp,Lohmann:2005im}, and similar constraints as the tau $g-2$ and EDM are obtained. 
Although it is expected that the WDM predictions are smaller than the tau $g-2$ and EDM due to the size of $Z$-scalar and $Z$-fermion couplings~\cite{Arroyo-Urena:2015uoa,Arroyo-Urena:2016ygo,Arroyo-Urena:2017sfb}, the recent study for $e^+ e^- \to Z/\gamma^* \to \tau^+ \tau^-$ in several collider experiments~\cite{Lu:2025heu} showed that the sensitivities for these dipole moments are significantly improved. 
Note that the analytical expressions for amplitudes of these processes can be found in, e.g., ref.~\cite{Gogniat:2025eom}, including one-loop radiative corrections. 
When we apply the current analyses to the off-shell photon and/or the WDMs, we may be able to find another footprint of the radiative mass models.

\section*{Acknowledgments}

We would like to thank Zhonglv Huang for helpful discussions. 
YN is supported by Natural Science Foundation of Shanghai. 
YS is supported by Natural Science Foundation of China under grant No. W2433006.

\appendix

\section{Oblique parameters}
\label{app:STU}

The Peskin-Takeuchi parameters $S, T, U$~\cite{STU_Peskin:1991sw}, also known as the oblique parameters, are usually used in testing electroweak precisions by modifying the propagators of the gauge fields. 
The definitions can be generally expressed as
\begin{align}
S = \frac{4 s_w^2 c_w^2}{\alpha} \bar{S} \, , \qquad T = \frac{1}{\alpha} \bar{T} \, , \qquad U = \frac{4 s_w^2}{\alpha} \bar{U} \, ,
\label{eq:STU}
\end{align}
where
\begin{align}
\bar{S} &\equiv \frac{\Pi_{ZZ} (m^2_Z) - \Pi_{ZZ} (0)}{m^2_Z} - \frac{c_w^2 - s_w^2}{c_w s_w} {\left. \frac{\partial \Pi_{Z \gamma} (q^2)}{\partial q^2} \right|_{q^2 = 0} - \left. \frac{\partial \Pi_{\gamma \gamma} (q^2)}{\partial q^2} \right|_{q^2 = 0}} \, , \label{eq:Sbardef} \\[1ex]
\bar{T} &\equiv \frac{\Pi_{WW} (0)}{m^2_W} - \frac{\Pi_{ZZ} (0)}{m^2_Z} \, , \label{eq:Tbardef} \\[1ex]
\bar{U} &\equiv \frac{\Pi_{WW} (m^2_W) - \Pi_{WW} (0)}{m^2_W} - \frac{c_w}{s_w} {\left. \frac{\partial \Pi_{Z \gamma} (q^2)}{\partial q^2} \right|_{q^2 = 0} - \left. \frac{\partial \Pi_{\gamma \gamma} (q^2)}{\partial q^2} \right|_{q^2 = 0}} - c_w^2 \bar{S} \, . \label{eq:Ubardef}
\end{align}
Here, $\Pi_{V V'} (q^2)$ is the coefficient of $g^{\mu \nu}$ in the vacuum polarization tensor for $V V' = W W, Z Z, Z \gamma, \gamma \gamma$ with $q$ being the four-momentum of the gauge boson. 
For the radiative mass models in the present work, we have two kinds of contributions from exotic particles: the contribution of exotic scalars $\phi, \eta$ and that of exotic fermions $\psi_{L, R}$. 
Therefore, each parameter can be divided into
\begin{align}
S = S_s + S_f \, , \qquad T = T_s + T_f \, , \qquad U = U_s + U_f \, ,
\end{align}
where subscriptions show the scalar contributions ($s$) and the fermion contributions ($f$). 
Note that there will be some divergent parts in $S_{s, f}$, $T_{s, f}$ and $U_{s, f}$, but they are canceled with each other individually. 

The $W$ boson mass is modified by the $S, T, U$ parameters from the SM prediction as
\begin{align}
m_{W, {\rm pred}}^2 = m_{W, {\rm SM}}^2 \left[ 1 + \frac{\alpha}{c_w^2 - s_w^2} \left( - \frac{1}{2} S + c_w^2 T + \frac{c_w^2 - s_w^2}{4 s_w^2} U \right) \right] \, .
\label{eq:mW2pred}
\end{align}
For the general forms of $S_{s, f}, T_{s, f}, U_{s, f}$ are studied and summarized in refs.~\cite{Grimus:2007if,Grimus:2008nb,Albergaria:2023nby}. 
Note that we focus on the contributions at the one-loop level in the following discussion as well as in our numerical analyses.

\subsection{MF model}
\label{sec:STUinMF}

In the MF model, all exotic scalars contribute to the oblique parameters, while the exotic fermions do not, due to their $SU(2)_L \times U(1)_Y$ charges. 
Then, $S_f, T_f, U_f = 0$, and only scalar contributions are considered. 
In order to obtain these contributions, we need to diagonalize all mass matrices for neutral and charged scalars. 
Following refs.~\cite{Grimus:2007if,Grimus:2008nb}, the relevant diagonalizing matrices can be found as
\begin{align}
\mathcal{U} = \begin{pmatrix}
1 & 0 & 0 \\
0 & c_\theta & s_\theta
\end{pmatrix}\, , \qquad \mathcal{V} = \begin{pmatrix}
i & 1 & 0 & 0 \\
0 & 0 & 1 & i
\end{pmatrix} \, ,
\end{align}
and the mass eigenvalues are given by
\begin{align}
m_a^2 &= \{ 0, m_{\varphi_1^+}^2, m_{\varphi_2^+}^2 \} \, , \\[1ex]
\mu_b^2 &= \{ 0, m_{h^0}^2, m_{\sigma_\phi}^2  ,m^2_{a_\phi} \} \, ,
\end{align}
where the only mixing comes from exotic charged scalars, as mentioned in ref.~\cite{Khaw:2022qxh}. 
The first components in $m_a^2$ and $\mu_b^2$ correspond to the masses for the Nambu-Goldstone modes, $G^{\pm}$ and $G^0$.

\subsection{RS model}
\label{sec:STUinRS}

In the RS model, all exotic particles contribute to the oblique parameters. 
The scalar contributions can be obtained by the same way as in the MF model, and the relevant diagonalizing matrices are
\begin{align}
\mathcal{U} &= \begin{pmatrix}
1 & 0 \\
0 & 1
\end{pmatrix} \, , \qquad \mathcal{V} = \begin{pmatrix}
i & 1 & 0 & 0 & 0 \\
0 & 0 & R_{11} + i R_{31} & R_{12} + i R_{32} & R_{13} + i R_{33}
\end{pmatrix} \, , \label{eq:UVmatRS}
\end{align}
and the mass eigenvalues are given by
\begin{align}
m_a^2 &= \{ 0, m_{\phi^{\pm}}^2 \} \, , \\[1ex]
\mu_b^2 &= \{ 0, m_{h^0}^2, m_{\varphi_1}^2, m_{\varphi_2}^2, m_{\varphi_3}^2 \} \, . 
\end{align}
The fermion contributions can be calculated by following ref.~\cite{Albergaria:2023nby}, and the results are  obtained as
\begin{align}
S_f = - U_f = \frac{8 s_w^4}{\pi} \Bigl[ H(m_{\psi}^2, m_Z^2) - H(m_{\psi}^2, 0) \Bigr] \, , \qquad T_f = 0 \, ,
\end{align}
where we define
\begin{align}
H (M^2_1, M^2_2) = \int_0^1 \! dx \: x(1-x) \ln \left[ M^2_1 - M^2_2 x (1 - x) \right] \, .
\end{align}
The above results are consistent with those in ref.~\cite{Albergaria:2023nby}. 
The vanishing result of $T_f$ is coming from the dependence of the 4-momentum of the external $Z$ boson.

\section{$Z$-$\tau$-$\tau$ couplings}
\label{app:Ztautau-full}

Our exotic particles contribute to the $Z$-$\tau$-$\tau$ couplings at the one-loop level. 
The general couplings of the $Z$ boson to scalars and fermions can be derived from the kinetic terms, $|D_{\mu} \phi|^2$ and $i \bar{\psi} \gamma^{\mu} D_{\mu} \psi$, and are given by
\begin{align}
\text{scalar}:& ~~ \frac{g}{c_w} \cdot \phi^{\dagger} (T_3 - Q s_w^2) \overset{\leftrightarrow}{\partial_{\mu}} \phi \cdot Z^{\mu} \, , \\[1ex]
\text{fermion}:& ~~ \frac{g}{c_w} \cdot \bar{\psi} (T_3 - Q s_w^2) \gamma_{\mu} \psi \cdot Z^{\mu} \, ,
\end{align}
where $\phi^{\dagger} \overset{\leftrightarrow}{\partial_{\mu}} \phi \equiv \phi^{\dagger} (\partial_{\mu} \phi) - (\partial_{\mu} \phi^{\dagger}) \phi$. 
After applying our charge assignments and rotations, we can calculate the $Z$ boson couplings to our exotic particles in both models. 
Using these couplings, corrections to the $Z$-$\tau$-$\tau$ couplings defined in Eq.~\eqref{eq:LZtautau} are calculated via the on-shell renormalization scheme~\cite{Sirlin:1980nh,Hollik:1988ii}.

\subsection{MF model}
\label{sec:MFZtautau}

The corrections to the $Z$-$\tau$-$\tau$ couplings in the MF model can be obtained as
\begin{align}
\delta g_L^{\tau} &= - \frac{y_{\phi}^2 s_{2 \theta}^2}{32 \pi^2} \left( g_Z^{\eta} - g_Z^{\phi} \right) \Bigl[ s_{\alpha}^2 \delta F_1^{\rm (MF)} (m_Z^2) + c_{\alpha}^2 \delta F_2^{\rm (MF)} (m_Z^2) \Bigr] - \frac{y_{\phi} y_{\eta} s_{2 \theta} s_{2 \alpha}}{64 \pi^2} g_L^{\tau} \delta M^{\rm (MF)} \nonumber \\[0.5ex]
&\hspace{1.2em} + \frac{y_{\phi}^2}{32 \pi^2} \Bigl[ c_{\theta}^2 s_{\alpha}^2 \delta \widetilde{G}_{L \, (1, 1)}^{\rm (MF)} (m_Z^2) + s_{\theta}^2 s_{\alpha}^2 \delta \widetilde{G}_{L \, (1, 2)}^{\rm (MF)} (m_Z^2) \nonumber \\[0.3ex]
&\hspace{8.4em} + c_{\theta}^2 c_{\alpha}^2 \delta \widetilde{G}_{L \, (2, 1)}^{\rm (MF)} (m_Z^2) + s_{\theta}^2 c_{\alpha}^2 \delta \widetilde{G}_{L \, (2, 2)}^{\rm (MF)} (m_Z^2) \Bigr] \nonumber \\[0.5ex]
&\hspace{1.2em} + \frac{y_{\eta}^2}{32 \pi^2} g_L^{\tau} \Bigl[ s_{\theta}^2 c_{\alpha}^2 \delta G_{(1, 1)}^{\rm (MF)} + c_{\theta}^2 c_{\alpha}^2 \delta G_{(1, 2)}^{\rm (MF)}+ s_{\theta}^2 s_{\alpha}^2 \delta G_{(2, 1)}^{\rm (MF)} + c_{\theta}^2 s_{\alpha}^2 \delta G_{(2, 2)}^{\rm (MF)} \Bigr] \, , \label{eq:delgLMF} \\[1.0ex]
\delta g_R^{\tau} &= + \frac{y_{\eta}^2 s_{2 \theta}^2}{32 \pi^2} \left( g_Z^{\eta} - g_Z^{\phi} \right) \Bigl[ c_{\alpha}^2 \delta F_1^{\rm (MF)} (m_Z^2) + s_{\alpha}^2 \delta F_2^{\rm (MF)} (m_Z^2) \Bigr] - \frac{y_{\phi} y_{\eta} s_{2 \theta} s_{2 \alpha}}{64 \pi^2} g_R^{\tau} \delta M^{\rm (MF)} \nonumber \\[0.5ex]
&\hspace{1.2em} + \frac{y_{\eta}^2}{32 \pi^2} \Bigl[ s_{\theta}^2 c_{\alpha}^2 \delta \widetilde{G}_{R \, (1, 1)}^{\rm (MF)} (m_Z^2) + c_{\theta}^2 c_{\alpha}^2 \delta \widetilde{G}_{R \, (1, 2)}^{\rm (MF)} (m_Z^2) \nonumber \\[0.3ex]
&\hspace{8.4em} + s_{\theta}^2 s_{\alpha}^2 \delta \widetilde{G}_{R \, (2, 1)}^{\rm (MF)} (m_Z^2) + c_{\theta}^2 s_{\alpha}^2 \delta \widetilde{G}_{R \, (2, 2)}^{\rm (MF)} (m_Z^2) \Bigr] \nonumber \\[0.5ex]
&\hspace{1.2em} + \frac{y_{\phi}^2}{32 \pi^2} g_R^{\tau} \Bigl[ c_{\theta}^2 s_{\alpha}^2 \delta G_{(1, 1)}^{\rm (MF)} + s_{\theta}^2 s_{\alpha}^2 \delta G_{(1, 2)}^{\rm (MF)} + c_{\theta}^2 c_{\alpha}^2 \delta G_{(2, 1)}^{\rm (MF)} + s_{\theta}^2 c_{\alpha}^2 \delta G_{(2, 2)}^{\rm (MF)} \Bigr] \, , \label{eq:delgRMF}
\end{align}
where $g_Z^{\phi} = 1/2 - s_w^2$, $g_Z^{\eta} = - s_w^2$, and each function is defined by
\begin{align}
\delta F_a^{\rm (MF)} (q^2) &\equiv \sum_{j, k = 1}^2 (-1)^{j + k} C_{00} (m_{\tau}^2, q^2, m_{\tau}^2, m_{\psi_a}^2, m_{\varphi_j^+}^2, m_{\varphi_k^+}^2) \, , \label{eq:delFMF} \\[0.5ex]
\delta M^{\rm (MF)} &\equiv 2 m_{\tau} \cos \theta_{\rm MF} \Bigl[ \Bigr. m_{\psi_1} \left( B'_0 (m_{\tau}^2, m_{\psi_1}^2, m_{\varphi_1^+}^2) - B'_0 (m_{\tau}^2, m_{\psi_1}^2, m_{\varphi_2^+}^2) \right) \nonumber \\
&\hspace{7.0em} - m_{\psi_2} \left( B'_0 (m_{\tau}^2, m_{\psi_2}^2, m_{\varphi_1^+}^2) - B'_0 (m_{\tau}^2, m_{\psi_2}^2, m_{\varphi_2^+}^2) \right) \Bigl. \Bigr] \, , \label{eq:delMMF} \\[0.5ex]
\delta G_{(a, j)}^{\rm (MF)} &\equiv \frac{m_{\psi_a}^2 - m_{\varphi_j^+}^2}{m_{\tau}^2} \left( B_0 (m_{\tau}^2, m_{\psi_a}^2, m_{\varphi_j^+}^2) - B_0 (0, m_{\psi_a}^2, m_{\varphi_j^+}^2) \right) \nonumber \\[0.3ex]
&\hspace{10.0em} - \bigl( m_{\tau}^2 + m_{\psi_a}^2 - m_{\varphi_j^+}^2 \bigr) B'_0 (m_{\tau}^2, m_{\psi_a}^2, m_{\varphi_j^+}^2) \, , \label{eq:delGMF} \\[0.5ex]
\delta \widetilde{G}_{L(R) \, (a, j)}^{\rm (MF)} (q^2) &\equiv g_{L(R)}^{\tau} \delta G_{(a, j)}^{\rm (MF)} + 2 g_{L(R)}^{\tau} B_1 (m_{\tau}^2, m_{\psi_a}^2, m_{\varphi_j^+}^2) \nonumber \\[0.3ex]
&\hspace{4.0em} - 4 g_Z^{\phi(\eta)} C_{00} (m_{\tau}^2, q^2, m_{\tau}^2, m_{\psi_a}^2, m_{\varphi_j^+}^2, m_{\varphi_j^+}^2) \, , \label{eq:delGtilMF}
\end{align}
with relevant Passarino-Veltman integrals defined in appendix~\ref{app:PVint}. 
Note that due to $g_L^{\phi} = - g_Z^{\tau}$ and $g_Z^{\eta} = - g_R^{\tau}$, all the above functions do not have UV divergent parts. 
Moreover, in the limit of $m_{\tau}^2 / m_{\psi_a}^2, m_Z^2 / m_{\psi_a}^2 \ll 1$, $\delta M^{\rm (MF)}$, $\delta G_{(a, j)}^{\rm (MF)}$ and $\delta \widetilde{G}_{L(R) \, (a, j)}^{\rm (MF)} (m_Z^2)$ give sub-leading contributions, and the dominant part of $\delta F_a^{\rm (MF)} (m_Z^2)$ is
\begin{align}
\delta F_a^{\rm (MF)} (m_Z^2) &= \frac{\bigl( x_{1, a}^2 + x_{2, a} (x_{1, a} - 2) \bigr) x_{1, a}}{4 (x_{1, a} - 1)^2 (x_{1, a} - x_{2, a})} \ln x_{1, a} - \frac{\bigl( x_{2, a}^2 + x_{1, a} (x_{2, a} - 2) \bigr) x_{2, a}}{4 (x_{2, a} - 1)^2 (x_{1, a} - x_{2, a})} \ln x_{2, a} \nonumber \\[0.5ex]
&\hspace{5.4em} - \frac{1}{2} - \frac{1}{4 (x_{1, a} - 1)} - \frac{1}{4 (x_{2, a} - 1)} + \mathcal{O} \biggl( \frac{m_{\tau}^2}{m_{\psi_a}^2}, \frac{m_Z^2}{m_{\psi_a}^2} \biggr) \, .
\end{align}
The leading part of this result is consistent with that in ref.~\cite{Baker:2020vkh}.

\subsection{RS model}
\label{sec:RSZtautau}

The corrections to the $Z$-$\tau$-$\tau$ couplings in the RS model can be obtained as
\begin{align}
\delta g_L^{\tau} &= - \frac{y_{\phi}^2}{32 \pi^2} g_Z^{\phi} \Bigl[ \left( 1 - R_{21}^2 \right) \delta F_{1, 2, 3}^{\rm (RS)} (m_Z^2) + \left( 1 - R_{22}^2 \right) \delta F_{2, 3, 1}^{\rm (RS)} (m_Z^2) + \left( 1 - R_{23}^2 \right) \delta F_{3, 1, 2}^{\rm (RS)} (m_Z^2) \Bigr] \nonumber \\[0.5ex]
&\hspace{1.2em} - \frac{y_{\phi} y_{\eta}}{8 \sqrt{2} \pi^2} \sum_{j = 1}^3 R_{1j} R_{2j} \delta M_{L \, j}^{\rm (RS)} (m_Z^2) \nonumber \\[0.5ex]
&\hspace{1.2em} + \frac{y_{\phi}^2}{64 \pi^2} \sum_{j = 1}^3 \left( 1 - R_{2j}^2 \right) \delta G_{(1), L \, j}^{\rm (RS)} (m_Z^2) + \frac{y_{\eta}^2}{32 \pi^2} \sum_{j = 1}^3 R_{2j}^2 \delta G_{(2), L \, j}^{\rm (RS)} (m_Z^2) \, , \label{eq:delgLRS} \\[1.0ex]
\delta g_R^{\tau} &= - \frac{y_{\phi} y_{\eta}}{8 \sqrt{2} \pi^2} \sum_{j = 1}^3 R_{1j} R_{2j} \delta M_{R \, j}^{\rm (RS)} (m_Z^2) \nonumber \\[0.5ex]
&\hspace{1.2em} + \frac{y_{\eta}^2}{32 \pi^2} \sum_{j = 1}^3 R_{2j}^2 \delta G_{(1), R \, j}^{\rm (RS)} (m_Z^2) + \frac{y_{\phi}^2}{64 \pi^2} \sum_{j = 1}^3 \left( 1 - R_{2j}^2 \right) \delta G_{(2), R \, j}^{\rm (RS)} (m_Z^2) \, , \label{eq:delgRRS}
\end{align}
where $g_Z^{\phi} = 1/2$, and each function is defined by
\begin{align}
\delta F_{j_1, j_2, j_3}^{\rm (RS)} (q^2) &\equiv B_1 (m_{\tau}^2, m_{\psi}^2, m_{\varphi_{j_1}}^2) - 2 C_{00} (m_{\tau}^2, q^2, m_{\tau}^2, m_{\psi}^2, m_{\varphi_{j_2}}^2, m_{\varphi_{j_3}}^2) \nonumber \\[0.3ex]
&\hspace{1.2em} + 2 C_{00} (m_{\tau}^2, q^2, m_{\tau}^2, m_{\psi}^2, m_{\varphi_{j_1}}^2, m_{\varphi_{j_2}}^2) + 2 C_{00} (m_{\tau}^2, q^2, m_{\tau}^2, m_{\psi}^2, m_{\varphi_{j_1}}^2, m_{\varphi_{j_3}}^2) \, , \label{eq:delFRS} \\
\delta M_{L(R) \, j}^{\rm (RS)} (q^2) &\equiv m_{\tau} m_{\psi} \left[ g_Z^{L(R)} B'_0 (m_{\tau}^2, m_{\psi}^2, m_{\varphi_j}^2) + g_Z^{\psi} C_0 (m_{\tau}^2, q^2, m_{\tau}^2, m_{\varphi_j}^2, m_{\psi}^2, m_{\psi}^2) \right] \, , \label{eq:delMRS} \\[0.5ex]
\delta G_{(1), L(R) \, j}^{\rm (RS)} (q^2) &\equiv g_{L(R)}^{\tau} \delta \widetilde{G}_{(1) \, j}^{\rm (RS)} + 2 g_Z^{\psi} \left[ \delta \widetilde{G}_{(2) \, j}^{\rm (RS)} (q^2) + 2 m_{\tau}^2 C_1 (m_{\tau}^2, q^2, m_{\tau}^2, m_{\varphi_j}^2, m_{\psi}^2, m_{\psi}^2) \right] \, , \label{eq:delG1RS} \\[0.5ex]
\delta G_{(2), L(R) \, j}^{\rm (RS)} (q^2) &\equiv g_{L(R)}^{\tau} \delta \widetilde{G}_{(1) \, j}^{\rm (RS)} - 2 g_Z^{\psi} m_{\tau}^2 \Bigl[ \Bigr. C_0 (m_{\tau}^2, q^2, m_{\tau}^2, m_{\varphi_j}^2, m_{\psi}^2, m_{\psi}^2) \nonumber \\[0.3ex]
&\hspace{12.0em} + 2 C_1 (m_{\tau}^2, q^2, m_{\tau}^2, m_{\varphi_j}^2, m_{\psi}^2, m_{\psi}^2) \Bigl. \Bigr] \, , \label{eq:delG2RS}
\end{align}
with $g_Z^{\psi} = + s_w^2$ and auxiliary functions $\delta \widetilde{G}_{(1, 2), L(R) \, j}^{\rm (RS)} (q^2)$ begin defined as
\begin{align}
\delta \widetilde{G}_{(1) \, j}^{\rm (RS)} &= \frac{m_{\psi}^2 - m_{\varphi_j}^2}{m_{\tau}^2} \left( B_0 (m_{\tau}^2, m_{\psi}^2, m_{\varphi_j}^2) - B_0 (0, m_{\psi}^2, m_{\varphi_j}^2) \right) \nonumber \\[0.3ex]
&\hspace{10.0em} - \bigl( m_{\tau}^2 + m_{\psi}^2 - m_{\varphi_j}^2 \bigr) B'_0 (m_{\tau}^2, m_{\psi}^2, m_{\varphi_j}^2) \, , \label{eq:delG1tilRS} \\[0.5ex]
\delta \widetilde{G}_{(2) \, j}^{\rm (RS)} (q^2) &= B_1 (m_{\tau}^2, m_{\psi}^2, m_{\varphi_j}^2) + B_0 (q^2, m_{\psi}^2, m_{\psi}^2) - 2 C_{00} (m_{\tau}^2, q^2, m_{\tau}^2, m_{\varphi_j}^2, m_{\psi}^2, m_{\psi}^2) \nonumber \\[0.3ex]
&\hspace{3.0em} - \bigl( m_{\psi}^2 - m_{\varphi_j}^2 \bigr) C_0 (m_{\tau}^2, q^2, m_{\tau}^2, m_{\varphi_j}^2, m_{\psi}^2, m_{\psi}^2) \, . \label{eq:delG2tilRS}
\end{align}
In the limit of $m_{\tau}^2 / m_{\psi_a}^2, m_Z^2 / m_{\psi_a}^2 \ll 1$, $\delta F_{j_1, j_2, j_3}^{\rm (RS)} (m_Z^2)$ give leading contributions, while others are sub-leading ones, and therefore, $\delta g_L^{\tau}$ is relevant to the corrections to the $Z$-$\tau$-$\tau$ couplings. 
Note that the dominant part can be expressed as
\begin{align}
\delta F_{j_1, j_2, j_3}^{\rm (RS)} (m_Z^2) &= \frac{x_{j_1} (x_{j_1} - 1) + x_{j_1} (x_{j_1} - 2) \ln x_{j_1}}{2 (x_{j_1} - 1)^2} + \frac{1}{2 (x_{j_2} - x_{j_3})} \left( \frac{x_{j_2}^2 \ln x_{j_2}}{x_{j_2} - 1} - \frac{x_{j_3}^2 \ln x_{j_3}}{x_{j_3} - 1} \right) \nonumber \\[0.3ex]
&\hspace{1.2em} - \sum_{k = j_2, j_3} \frac{1}{2 (x_{j_1} - x_k)} \left( \frac{x_{j_1}^2 \ln x_{j_1}}{x_{j_1} - 1} - \frac{x_k^2 \ln x_k}{x_k - 1} \right) + \mathcal{O} \biggl( \frac{m_{\tau}^2}{m_{\psi}^2}, \frac{m_Z^2}{m_{\psi}^2} \biggr) \, .
\end{align}

\section{Passarino-Veltman integrals}
\label{app:PVint}

We here summarize Passarino-Veltman integrals~\cite{Passarino:1978jh} relevant for our numerical analyses. 
The self-energy type integrals are
\begin{align}
B_0 (p^2, m_0^2, m_1^2) &= \Delta_{\epsilon} - \int_0^1 \! dx \ln \left[ \frac{- x (1 - x) p^2 + x m_1^2 + (1 - x) m_0^2}{\mu^2} \right] \, , \\[1ex]
B'_0 (p^2, m_0^2, m_1^2) &= \frac{\partial}{\partial p^2} B_0 (p^2, m_0^2, m_1^2) = \int_0^1 \! dx \frac{x (1 - x)}{- x (1 - x) p^2 + x m_1^2 + (1 - x) m_0^2} \, , \\[1ex]
B_1 (p^2, m_0^2, m_1^2) &= - \frac{1}{2} \Delta_{\epsilon} + \int_0^1 \! dx \: x \ln \left[ \frac{- x (1 - x) p^2 + x m_1^2 + (1 - x) m_0^2}{\mu^2} \right] \, ,
\end{align}
and the triangle type integrals are
\begin{align}
C_0 (p_1^2, q^2, p_2^2, m_0^2, m_1^2, m_2^2) &= - \int_0^1 \! d x_1 \int_0^{1 - x_1} \! d x_2 \frac{1}{C_3} \, , \\[1ex]
C_{00} (p_1^2, q^2, p_2^2, m_0^2, m_1^2, m_2^2) &= \frac{1}{4} \Delta_{\epsilon} - \frac{1}{2} \int_0^1 \! d x_1 \int_0^{1 - x_1} \! d x_2 \ln \left[ \frac{C_3}{\mu^2} \right] \, , \\[1ex]
C_1 (p_1^2, q^2, p_2^2, m_0^2, m_1^2, m_2^2) &= \int_0^1 \! d x_1 \int_0^{1 - x_1} \! d x_2 \frac{x_1}{C_3} \, ,
\end{align}
with $C_3 = x_1 (x_1 + x_2 - 1) p_1^2 + x_2 (x_1 + x_2 - 1) p_2^2 - x_1 x_2 q^2 + x_1 m_1^2 + x_2 m_2^2 + (1 - x_1 - x_2) m_0^2$ and $q^2 = (p_2 - p_1)^2$. 
Here, $\Delta_{\epsilon}$ indicates the UV divergent part which is defined as
\begin{align}
\Delta_{\epsilon} \equiv \frac{2}{\epsilon} - \gamma_E + \ln 4 \pi \, ,
\end{align}
with $\epsilon = 4 - D$ and the Euler-Mascheroni constant $\gamma_E \approx 0.5772$. 
Note that when $q^2 = 0$ with $p_1^2 = p_2^2$ and $m_1^2 = m_2^2$, $C_0$ and $C_1$ functions can be simplified as
\begin{align}
C_{0, 1} (p_1^2, 0, p_1^2, m_0^2, m_1^2, m_1^2) &= \frac{1}{m_0^2} \hat{C}_{0, 1} (p_1^2, m_0^2, m_1^2) = \frac{1}{m_1^2} \widetilde{C}_{0, 1} (p_1^2, m_0^2, m_1^2) \, ,
\end{align}
where $\hat{C}_{0, 1} (p_1^2, m_0^2, m_1^2)$ and $\widetilde{C}_{0, 1} (p_1^2, m_0^2, m_1^2)$ are dimensionless functions,
\begin{align}
\hat{C}_0 (p_1^2, m_0^2, m_1^2) &\equiv - \int_0^1 \! d x \frac{(1 - x) m_0^2}{- x (1 - x) p_1^2 + x m_0^2 + (1 - x) m_1^2} = \frac{m_0^2}{m_1^2} \widetilde{C}_0 (p_1^2, m_0^2, m_1^2) \, , \\
\hat{C}_1 (p_1^2, m_0^2, m_1^2) &\equiv \frac{1}{2} \int_0^1 \! d x \frac{(1 - x)^2 m_0^2}{- x (1 - x) p_1^2 + x m_0^2 + (1 - x) m_1^2} = \frac{m_0^2}{m_1^2} \widetilde{C}_1 (p_1^2, m_0^2, m_1^2) \, .
\end{align}

\bibliographystyle{utphys}
\bibliography{tauDipole}

\end{document}